\documentclass[twocolumn,prl,floatfix,longbibliography,showpacs,superscriptaddress]{revtex4-2}

\usepackage[caption=false]{subfig} 
\usepackage{ragged2e} 
\DeclareCaptionJustification{justified}{\justifying}

\usepackage{bm} 
\usepackage{placeins}
\usepackage{todonotes}
\usepackage{subfig}
\usepackage{amsmath}
\usepackage{amsfonts}
\usepackage{amsbsy}
\usepackage{xcolor}
\usepackage{soul}
\usepackage{graphicx}
\usepackage{epsfig}
\usepackage{hyperref}
\usepackage{txfonts}
\usepackage{mathtools}
\usepackage{multirow}
\usepackage{dsfont}
\usepackage{balance} 
\hypersetup{
    unicode=true, 
    plainpages=false,
    colorlinks=true,
    linkcolor=blue,
    citecolor=blue,
    filecolor=blue,
    urlcolor=blue
}
\usepackage{bbold}
\usepackage{amsmath}
\usepackage{amsfonts}
\usepackage{amsbsy}
\usepackage{xcolor}
\usepackage{soul}
\usepackage{graphicx}
\usepackage{epsfig}

\newcommand{\pol}{\hat{\bf e}}
\newcommand{\rv}{{\bf r}}

\newcommand{\qv}{{\bf q}}
\newcommand{\kv}{{\bf k}}

\newcommand{\eo}{\epsilon_0}
\newcommand{\beq}{\begin{equation}}
\newcommand{\eeq}{\end{equation}}
\newcommand{\bea}{\begin{eqnarray}}
\newcommand{\eea}{\end{eqnarray}}
\newcommand{\BEQAL}{\begin{align}}
\newcommand{\EEQAL}{\end{align}}

\newcommand{\comment}[1]{{}}

\newcommand{\commentout}[1]{{}}

\newcommand{\radKernel}{\ensuremath{\mathsf{G}}}

\makeatletter
\renewcommand{\fnum@figure}{\textbf{Fig.} \thefigure}
\makeatother

\renewcommand{\thefigure}{\textbf{\arabic{figure}}}

\makeatletter
\def\@hangfrom@section#1#2#3{\@hangfrom{#1#2}#3}
\def\@hangfroms@section#1#2{#1#2}
\makeatother

\begin{document}
\title{Emergence of an Epsilon-Near-Zero Medium from Microscopic Atomic Principles}
\author{L. Ruks}
\affiliation{NTT Basic Research Laboratories \& NTT Research Center for Theoretical Quantum Information, NTT, Inc., 3-1 Morinosato Wakamiya, Atsugi, Kanagawa, 243-0198, Japan}
\author{J. Ruostekoski}
\affiliation{Department of Physics, Lancaster University, Lancaster, LA1 4YB, United Kingdom}

\begin{abstract}
We demonstrate that an effective near-zero refractive index can emerge from collective light scattering in a discrete atomic lattice, using essentially exact microscopic simulations. In a 25-layer array, cooperative response leads to over a thirtyfold increase in the effective optical wavelength within the medium, almost eliminating optical phase accumulation, with potential applications in spectroscopy and optical manipulation of quantum emitters. Crucially, the near-zero refractive index arises from first-principles microscopic theory, rather than being imposed through continuous phenomenological effective-medium models---providing conceptually important insight into the emergence of macroscopic electromagnetism from atomic-scale interactions.
\end{abstract}

\date{\today}
\maketitle

Effective-medium theories form the foundation of standard macroscopic electrodynamics~\cite{Jackson,BOR99}, describing quantities such as refractive index and permittivity by averaging over microscopic constituents. However, these models inherently neglect detailed information on position-dependent dipole-dipole interactions and assume that each scatterer responds only to the average field of its surroundings. Understanding how macroscopic electromagnetic behavior emerges from microscopic atomic principles in granular systems of discrete scatterers remains a fundamentally nontrivial problem~\cite{Javanainen2014a,JavanainenMFT,Jenkins_thermshift,Andreoli21}.

Among the extraordinary predictions of macroscopic electrodynamics of continuous media is the epsilon-near-zero (ENZ) regime, where the material permittivity and refractive index approach zero~\cite{Liberal17,Lobet23,Xie2025}. ENZ media exhibit striking features such as diverging optical wavelength and phase velocity, leading to a decoupling of spatial and temporal field variations and uniform phase across the medium. These effects underpin a range of applications in wavefront shaping~\cite{Enoch02,Alu07,Pacheco14, Soric15},
magnification and geometry-insensitive guiding~\cite{Silveirinha06,Edwards08,Adams11}, enhanced optical nonlinearities~\cite{Argyropoulos12,Zahirul16,Reshef19,Yang19}, and tailored light–matter interactions~\cite{Liberal17c,Campione15b}. The ENZ regime emerges in electric conductors at the reduced plasma frequency~\cite{kinsey19}, and is explored in metamaterials~\cite{Vesseur13,Mass13,Li15}
and photonic crystals~\cite{Huang11,Chan12}.
While ENZ behavior in these systems have been introduced via phenomenological models or metamaterial design, its emergence from microscopic atomic principles remains unexplored. 

Here, we perform microscopic simulations of light propagation in atomic arrays and demonstrate the emergence of ENZ behavior arising from collective light scattering. 
In the low light intensity (LLI) limit, our atomistic simulations are exact for stationary atoms~\cite{Javanainen1999a,Lee16} and
this regime has been shown in experiments to accurately describe optical responses even in dense atomic ensembles~\cite{Rui2020,Vatre2024}.
We observe more than a thirtyfold increase in the effective optical wavelength near collective resonances at the band edge of the lattice dispersion relation, leading to a pronounced suppression of phase accumulation across the medium that is robust to atomic position fluctuations. Our results apply broadly to point-dipole scatterers in different physical systems and provide an atomistic description of ENZ optical response, offering new insight into the microscopic origins of bulk electromagnetic behavior.

Atomic arrays are increasingly studied as platforms for cooperative light–matter interactions~\cite{Ruostekoski2023,Reitz22} and can be implemented in optical lattices with subwavelength spacing and high control. Lattices with unit filling can be prepared with hundreds of thousands of atoms~\cite{Shao2024}, and transmission experiments have observed spectral resonance narrowing below the fundamental quantum limit, set by the single-atom Wigner-Weisskopf linewidth~\cite{Rui2020}, and coherent optical switching~\cite{Srakaew22}. While early theoretical work on cooperatively responding arrays focused, e.g., on subradiance~\cite{Plankensteiner2015,Facchinetti16,Jen16,Bettles2015d,Sutherland1D,Asenjo-Garcia2017a}, more recent studies have also explored them as versatile platforms for enhanced light-matter coupling, including excitation mediation~\cite{Guimond2019,Ballantine20ant,Shah24}, cavity-like effects~\cite{Plankensteiner2017,Parmee20bistable,Pedersen23,Castells25,Sinha2025}, and coupling to long-wavelength Rydberg excitations~\cite{Bekenstein2020,Cardoner21,Zhang2022}.

\begin{figure}
    \centering
    \includegraphics[width=\linewidth]{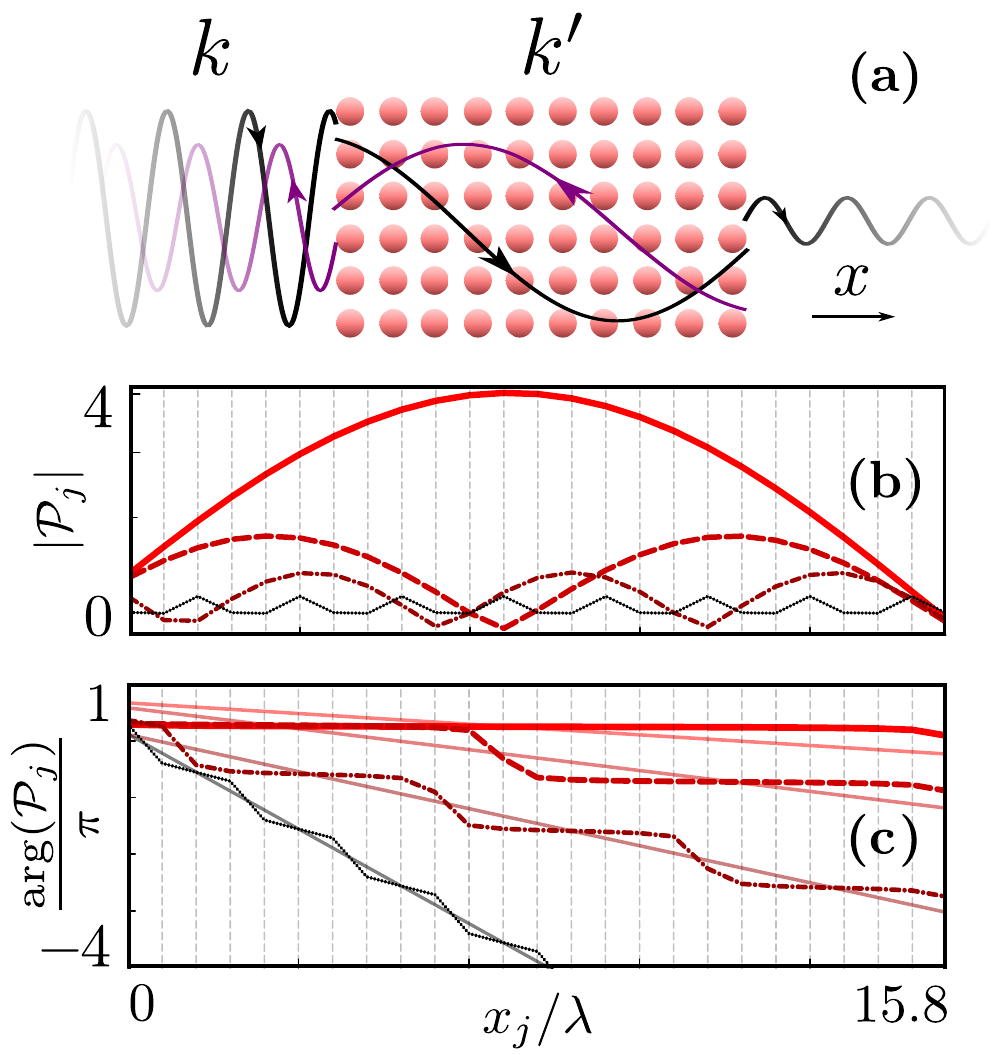}
    \caption{Light propagation in an effective ENZ atomic medium. (a) Schematic of light transmitted through a cubic atom array, infinitely extended in the $y$ and $z$ directions. The incident beam with wavevector $\kv=k\pol_x$ excites the atomic polarization, Eq.~\eqref{eq:planepol}, consisting of forwards and backwards 
propagating waves with wavevectors $\pm k'\pol_x$. (b) Magnitude $|\mathcal{P}_{j}|$ [in units of $\mathcal{D}\epsilon_{0}/(\hbar\gamma)$] and (c) phase $\mathrm{arg}(\mathcal{P}_{j})$ of atomic polarization layer amplitudes as a function of layer position $x_{j}$ (marked by vertical dashed lines). The detunings are $\Delta/\gamma = -0.025,0.103, 0.115, 0.119,$ corresponding to the black dotted, maroon dash-dotted, red dashed, and bright red solid lines, respectively, and the faint lines in (c) the corresponding backward wave components. From these we obtain medium wavenumbers $k'/k = 0.25, 0.1, 0.06, 0.03$. }
    \label{fig:fig1}
\end{figure}

We consider a cubic lattice of atoms, with spacing $a$ and the lattice vectors aligned along the Cartesian axes. To examine the large array limit in the $yz$ plane, we represent the system as a stack of $N_{x}$ identical planar arrays, located at $x=x_{\ell} = a\ell$ ($\ell=0,\ldots,N_x-1$), forming a slab of atoms.
The system is illuminated by a coherent laser field propagating in the positive $x$-direction, with the positive frequency component $\bm{\mathcal{E}}^{+}(\textbf{r}) = \pol_{\rm in} {\mathcal{E}} e^{i\kv\cdot \rv}$, dominant wavevector $\textbf{k} = (k_{x},\textbf{k}_{\parallel})$, and wavelength $\lambda=2\pi/k$, as depicted in Fig.~\ref{fig:fig1}(a).
Assuming atoms with a single electronic ground state and a degenerate excited-state manifold, we write the atomic polarization amplitude $\mathcal{D}\bm{\mathcal{P}}^{(\textbf{n})}$ for atom $\textbf{n}$ at position $\textbf{r}_{\textbf{n}}$, where $\mathcal{D}$ denotes the reduced dipole matrix element and the excitation amplitude $\mathcal{P}^{(\textbf{n})}$ spans the available polarization modes. We denote light and atomic excitation amplitudes using their slowly varying positive-frequency components, with rapid oscillations $e^{-i\omega t}$ at the laser frequency $\omega$ factored out~\cite{Ruostekoski1997a}. 
$\mathcal{P}^{(\textbf{n})}$ is driven by the incident Rabi frequency $\bm{\mathcal{R}}^{+}(\textbf{r}_{\textbf{n}})=\mathcal{D}\bm{\mathcal{E}}^{+}(\textbf{r}_{\textbf{n}})/\hbar$ and the field $\epsilon_0\textbf{E}_{\mathrm{s}}^{\textbf{n}'}(\textbf{r}) = \mathsf{G}(\textbf{r} - \textbf{r}_{\textbf{n}'})\mathcal{D}\mathcal{P}^{(\textbf{n}')}$ scattered by all other atoms $\textbf{n}'$, where $\mathsf{G}(\rv)$ is the free-space dipole radiation kernel~\cite{Jackson} (End Note). We assume the LLI limit, where the atoms behave as classical linearly coupled dipoles, and their evolution is governed by the coupled-dipole equations~\cite{Javanainen1999a,Lee16,Lax51,Morice1995a,Ruostekoski1997a,Sokolov2011},
\begin{align}\label{eq:point-scattering-main}
\dot{\bm{\mathcal{P}}}^{(\textbf{n})}
 =   \left( i \Delta - \gamma \right)
{\bm{\mathcal{P}}}^{(\textbf{n})} +   i\bm{\mathcal{R}}^{+}(\textbf{r}_{\textbf{n}})+i\xi\sum_{\textbf{n}' \neq \textbf{n}}\radKernel(\textbf{r}_{\textbf{n}} - \textbf{r}_{\textbf{n}'}){\bm{\mathcal{P}}}^{(\textbf{n}')},
\end{align}
where  $\xi=\mathcal{D}^2/(\hbar\eo)$, $\gamma$ is the single-atom linewidth, and $\Delta=\omega-\omega_0$ is the laser detuning from the atomic resonance frequency $\omega_{0}$. 

It is efficient to exploit the translational invariance in the $yz$ plane by reframing Eq.~\eqref{eq:point-scattering-main} in terms of the coupling between collective excitations of each layer~\cite{ruks25}, with well-defined in-plane quasimomentum $\qv_{\parallel}$. By transforming Eq.~\eqref{eq:point-scattering-main} into momentum space, the amplitudes $\bm{\mathcal{P}}_{\ell}(\qv_{\parallel})$ of these excitations satisfy the analogous evolution,
\begin{align}
\dot{\bm{{\mathcal{P}}}}_{\ell} = & \big[i\Delta-\gamma+\xi\sum_{j \neq 0}e^{i\qv_{\parallel}\cdot \mathbf{r}_{\parallel j}}\mathsf{G}(\textbf{r}_{\parallel j})\big] \bm{\mathcal{P}}_{\ell}+i{\bm{\mathcal{R}}}^{+}_{\ell} \nonumber\\
    &+ i\xi\sum_{m\neq \ell}^{N_{x}-1}\mathsf{G}^{\text{L}}\boldsymbol{(}(x_{\ell} - x_{m})\hat{\textbf{e}}_{x},\qv_{\parallel}\boldsymbol{)}\bm{\mathcal{P}}_{m},
    \label{eq:layer-layer-scattering}
\end{align}
for the in-plane Fourier component $\bm{\mathcal{R}}_{\ell}^{+}(\textbf{q}_{\parallel})$ of the Rabi frequency in the plane $x = x_{\ell}$. 
The sum in the first line describes the intralayer interaction, with the self-interaction term at $\textbf{r}_{\parallel j} = \textbf{0}$ excluded. 
The interaction between layers is captured in the layer propagator $\mathsf{G}^{\mathrm{L}}(\textbf{r},\qv_{\parallel}) =  \sum_{j} e^{i\qv_{\parallel}\cdot \mathbf{r}_{\parallel j}}\mathsf{G}(\textbf{r} - \textbf{r}_{\parallel j})$, which acts as the dipole radiation kernel for a layer located at $x = 0$ and excited with in-plane quasimomentum $\textbf{q}_{\parallel}$. $\mathsf{G}^{\mathrm{L}}$ can be expressed as a sum of plane waves with wavevector $\qv$, along with evanescent fields that contribute to the rapidly varying near-field structure (see End Matter). The field scattered from all of the layers is then similarly expressed:
\beq
\label{eq:scattered-layer}
\epsilon_0 {\textbf{E}}{}^+_s(\textbf{r}) = \epsilon_0\sum_{\textbf{n}}\textbf{E}_{\mathrm{s}}^{\textbf{n}}(\textbf{r}) = \sum_{\qv_{\parallel}}\sum_{\ell=0}^{N_{x}-1}\mathsf{G}^{\text{L}}(\textbf{r} - x_{\ell}\hat{\mathbf{e}}_{x})\mathcal{D}\bm{\mathcal{P}}_{\ell}(\qv_{\parallel}).
\eeq 

Using the general formalism for coupled stacked layers, we have demonstrated ENZ response across a broad class of point-dipole systems---dipoles with fixed orientation (as in two-level atoms in the LLI limit and standard dipolar resonator models), isotropic dipoles, and configurations with finite-width beams or different lattice constants. In the following, we focus on isotropic dipoles corresponding to a  $J = 0 \to J' = 1$ atomic transition, with lattice spacing $a = 0.66\lambda$ and $N_{x} = 25$ layers. We consider a normal incidence, $\kv=k\pol_x$, such that $\kv_{\parallel} = \textbf{0}$ and $\qv_{\parallel} =0$, with dipole amplitudes aligned with the light polarization and reduced to scalars, $\bm{\mathcal{P}}^{(\textbf{n})} = \hat{\textbf{e}}_{\mathrm{in}}{\mathcal{P}}^{(\textbf{n})}$. The phase reference is determined by setting ${\mathcal{E}}$ real.

To establish macroscopic electrodynamical behavior of light, we consider the following ansatz for the light field envelope inside the sample which consists of oppositely propagating plane waves, 
\beq
\mathbf{E}_m^{+}(\textbf{r}) = \hat{\textbf{e}}_{{\rm in}} {{\mathcal{E}}}_{f}e^{i k' x} + \hat{\textbf{e}}_{{\rm in}} {{\mathcal{E}}}_{b}e^{-i k' x},
\label{eq:planelight}
\eeq
from which the effective phase refractive index is defined as $n_p = k'/k$, for $\mathrm{Im}(k') \geq 0$. Due to the discrete nature of atomic positions, the total light field also includes a rapidly varying microscopic component with the periodicity of the lattice constant. These short-wavelength variations can be removed via coarse-graining, yielding smooth macroscopic electromagnetic fields.

To isolate this macroscopic behavior, we first extract wavevectors for the atomic polarization density amplitudes, obtained from the steady-state solutions of Eq.~\eqref{eq:layer-layer-scattering}, for the layer amplitudes $\bm{\mathcal{P}}_{\ell}$. When these amplitudes decompose into the forward and backward propagating plane waves, 
\beq
\bm{\mathcal{P}}_{\ell} = \hat{\textbf{e}}_{\mathrm{in}}{\mathcal{P}}_{\ell} =  \hat{\textbf{e}}_{\mathrm{in}}\mathcal{P}_{f}e^{i\ell k'a} + \hat{\textbf{e}}_{\mathrm{in}} \mathcal{P}_{b}e^{-i\ell k'a},
\label{eq:planepol}
\eeq
the medium light field envelope does likewise, sharing a common wavenumber $k'$. We show that this effective phase refractive index generically approaches zero when exciting ENZ collective resonances, whose effective wavenumber is dramatically reduced. 

\begin{figure}
	\centering
	\includegraphics[width=\linewidth]{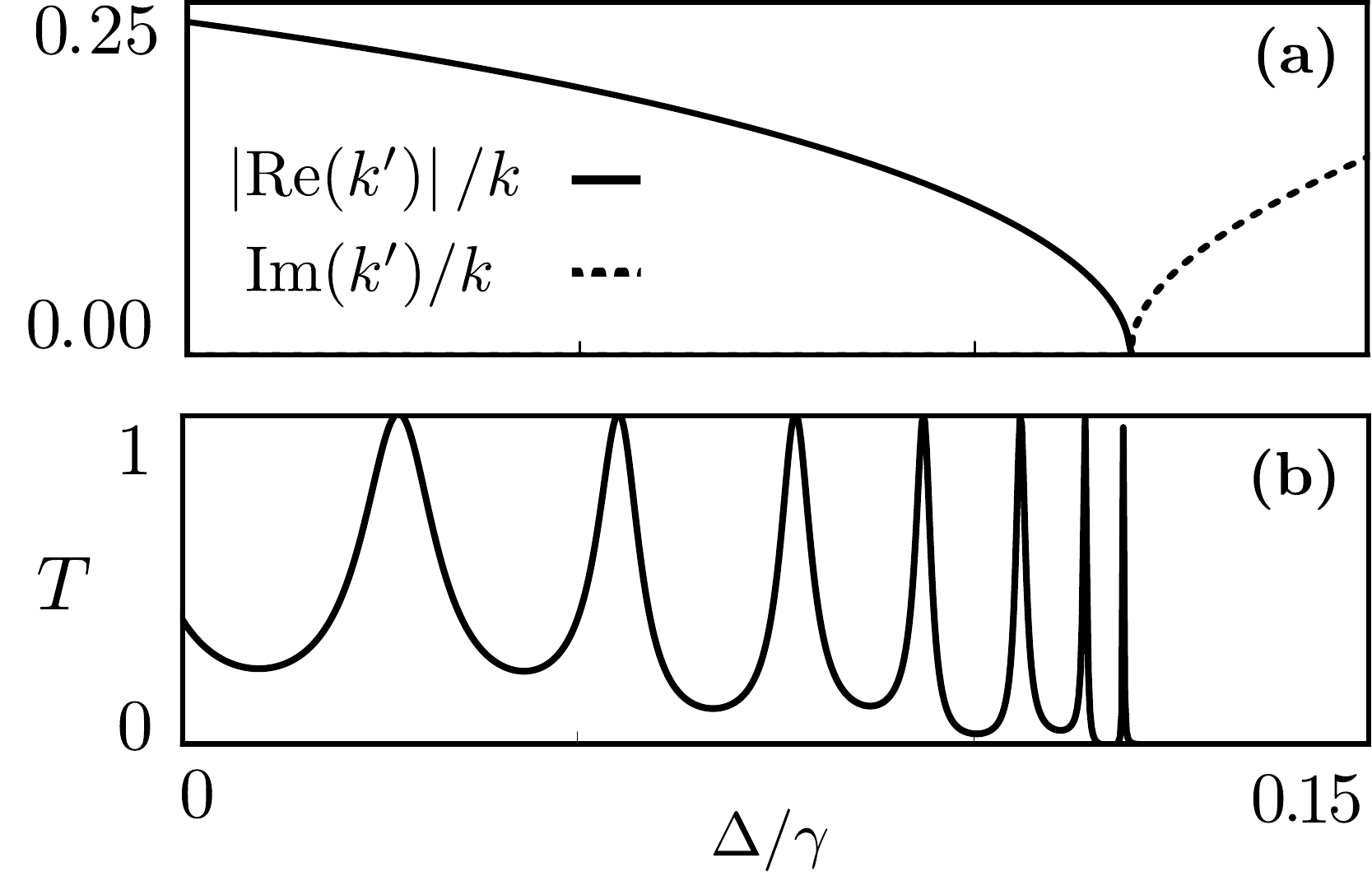}
	\caption{(a) Effective wavenumber $k'$ of atomic polarization within and (b) power transmission $T$ through 
    the medium as a function of laser detuning $\Delta$. 
    In (a), $k'$ is purely real for $\Delta\alt  0.12\gamma$ and imaginary for $\Delta\agt  0.12\gamma$. }
	\label{fig:fig2}
\end{figure}

In Fig.~\ref{fig:fig1}(b), we show a pronounced increase in the excitation wavelength of the atomic polarization within the medium as the laser detuning is varied. The excitation amplitudes are accurately fitted to the ansatz of oppositely propagating plane waves in Eq.~\eqref{eq:planepol}, with relative residual error of less than $10^{-2}$, yielding wavenumbers ranging from $k' = 0.25k$ down to $k' = 0.03k$---corresponding to an effective wavelength of approximately $33\lambda$. This significant reduction in wavenumber is illustrated in Fig.~\ref{fig:fig2}(a) where a purely real $k'$ decreases monotonically with detuning until the effective wavelength approaches twice the medium thickness at $\Delta \simeq 0.12\gamma.$ Beyond this point, the system transitions to an imaginary wavenumber, indicating exponentially decaying excitations within the medium. We observe similar behavior---fully consistent with Eq.~\eqref{eq:planepol}---across a range of $N_{x}$, down to as few as five layers and over varying lattice constants, after adjusting detuning (see End Matter).
The increasing wavelength is dramatically illustrated in the accumulated phase across the sample in Fig.~\ref{fig:fig1}(c), which remains almost constant throughout the medium in extreme cases. 

\begin{figure}
	\centering
	\includegraphics[width=\linewidth]{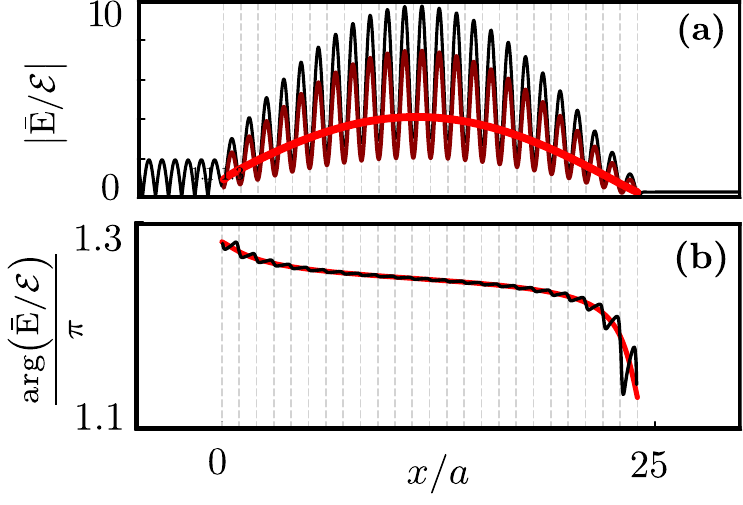}
	\caption{Microscopic light field on atomic scale. (a) Magnitude and (b) phase of the total microscopic light field (dotted-black line) along the incident field polarisation as the line $(y,z) = (a/2,a/2)$ is continuously traversed in between the atoms, $-5a \leq x \leq 30a$, for the detuning $\Delta/\gamma = 0.119$ giving polarization amplitudes presented in Fig.~\ref{fig:fig1}(b). Averaged amplitudes $\bar{\mathrm{E}}^{+}$ are obtained by convolution of the microscopic field with a Gaussian of the root-mean-square widths $\eta = 0.25a$ (dashed maroon line), and $\eta = a$ (solid bright red line), converging to the field amplitude Eq.~\eqref{eq:planelight}. 
    }
	\label{fig:fig3}
\end{figure}

The microscopic light field varies rapidly on the scale of the lattice constant due to the near-field dipole radiation (Fig.~\ref{fig:fig3}), as captured in the evanescent contributions to the scattered light.
However, when these nonpropagating components are ignored, the resulting light field envelope closely matches the slowly-varying atomic excitation profile, as described by Eq.~\eqref{eq:planelight}, with the same wavenumber $k'$. Remarkably, the phase remains almost uniform even in the presence of microscopic evanescent field components, with the phase variation more pronounced towards the far boundary.
The emergence of slowly varying light field is formally demonstrated in Fig.~\ref{fig:fig3}. Here, the microscopic field is convolved with Gaussians of varying root-mean-square width $\eta$, which acts a length over which the field is averaged. 
As $\eta$ increases from zero, short-range near-field variations are suppressed, and a smooth macroscopic field emerges once the averaging length exceeds the lattice constant, $\eta \gtrsim a$. 
As expected, the amplitude profile in Fig.~\ref{fig:fig3} precisely matches the atomic response in Figs.~\ref{fig:fig1}(b,c). 
By extracting the phase refractive index $n_p = k'/k \to 0$, these results then demonstrate, using essentially exact microscopic calculations, how ENZ medium emerges from collective scattering.

In Fig.~\ref{fig:fig2}(b), we show light transmission as a function of detuning. A series of increasingly narrow Fano resonances appear, arising from interference of collective excitation eigenmodes with narrow subradiant and broad superradiant resonances---a direct signature of the atomic medium's granular nature.
Peaks at full power transmission $T = 1$ coincide with medium wavenumbers $k' \simeq j\pi/h$ ($j = 1,\ldots,N_x - 1$), for medium thickness $h = a(N_x - 1)$.
Beyond $\Delta \simeq 0.12\gamma$ in Fig.~\ref{fig:fig2}(b),  the incident light no longer couples to propagating modes and is reflected, resulting in evanescently decaying fields. We have found that the effect of a finite incident angle $\theta$ is to shift the transmission peaks in Fig.~\ref{fig:fig2}(b) to smaller detunings and, away from Fano resonances, transmission is generally suppressed once $|\theta| \gtrsim |\arcsin n_p|$. However, the presence of long-wavelength excitations with small $k'/k$ remains unchanged.

\begin{figure}
    \centering
    \includegraphics[width=\linewidth]{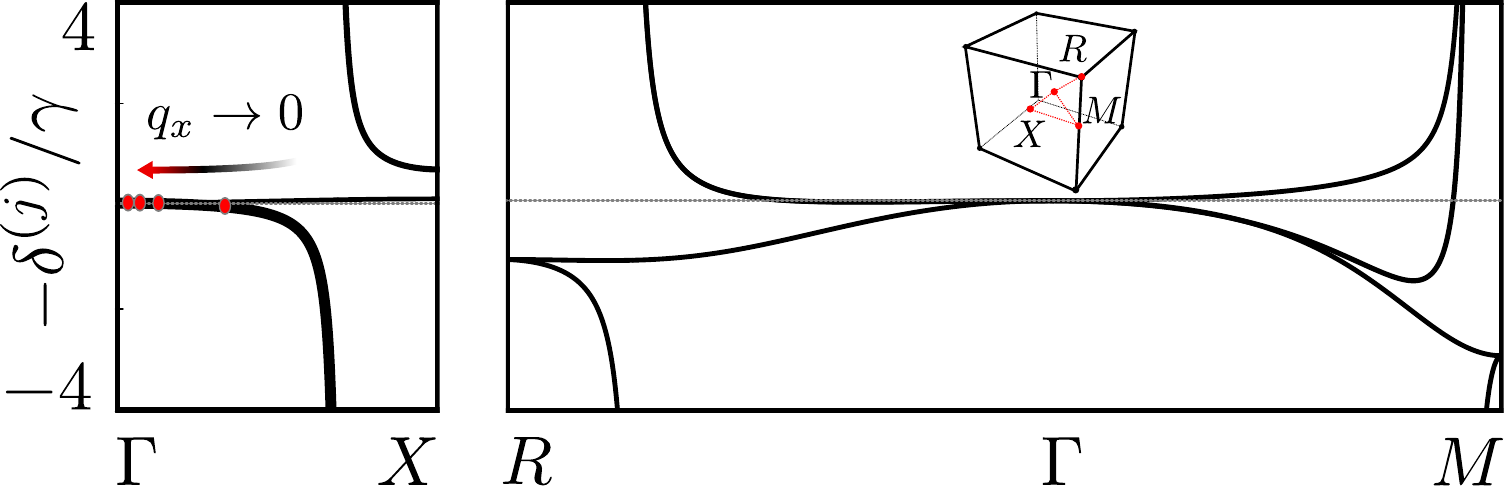}
    \caption{Collective line shifts $\delta^{(j)}$ of bands $j$ for an infinite lattice, presented for high-symmetry lines passing through the center of the 3D Brillouin zone (inset), where long-wavelength excitations are present. {The bold line denotes the collective resonance band, with quasimomenta $q_{y} = q_{z} = 0$, whose polarization lies in the $yz$ plane.} The dashed line denotes the edge of this band {at $q_{x} = 0$}, where $\delta \simeq -0.12\gamma,$ and the red markers correspond to the wavenumbers of resonant excitations extracted in Figs.~\ref{fig:fig1}(b-c).}
    \label{fig:fig4}
\end{figure}

We characterize the long-wavelength ENZ excitations by analyzing the collective resonances of an infinite atomic lattice ($N_x = \infty$). These resonances are obtained~\cite{castin09,Antezza09b} (see End Matter) by substituting the Bloch-wave ansatz $\mathbf{\mathcal{P}}_{\mathbf{n}} = \mathbf{\mathcal{P}}(\mathbf{q})e^{i\mathbf{q}\cdot\mathbf{r}_{\mathbf{n}}}$ into Eq.~\eqref{eq:point-scattering-main}, yielding a 3D linear system for $\mathbf{\mathcal{P}}(\mathbf{q})$ at each quasimomentum $\mathbf{q}$. The resulting eigensystem produces eigenvalues $\delta^{(j)}(\mathbf{q})$, which separate into distinct dispersion bands indexed by $j$ (Fig.~\ref{fig:fig4}), and represent the collective resonance frequency shifts from the single-atom resonance.

The excitation wavevectors $k'\pol_x$ of the forward-propagating waves and corresponding laser frequencies extracted from Fig.~\ref{fig:fig1} for the 25-layer lattice closely match these infinite-system collective resonances, satisfying $\Delta \simeq -\delta^{(j)}(\mathbf{q})$ and $k' \simeq q_x $ (with $q_y=q_z=0$), where $\qv$ is the resonant infinite-medium Bloch-wavevector, as demonstrated in Fig.~\ref{fig:fig4}. Additionally, an oppositely propagating wave is excited due to reflection from the far boundary of the lattice. The long-wavelength excitations with $k' \simeq 0$ observed in Fig.~\ref{fig:fig1}(b) occur when the laser is tuned to a collective resonance where $q_x \simeq 0$ and $\delta^{(j)}(\mathbf{q}) \simeq -0.12\gamma$. 
The diverging phase velocity $v_{p,x}=\omega/k' \rightarrow \infty$ at the resonance $\textbf{q} \simeq \textbf{0}$ in Fig.~\ref{fig:fig4} represents an increasing effective wavelength and emerging near-zero phase refractive index in Fig.~\ref{fig:fig2}(a). The band edge at $\delta^{(j)}(\mathbf{q}) \simeq -0.12\gamma$, on the other hand, gives rise to the transmission in Fig.~\ref{fig:fig2}(b), and is consistent with observations of an ENZ response in 1D continuous-media photonic crystals~\cite{Popov2019}.
Crucially, collective resonances at $\mathbf{q} = \mathbf{0}$ commonly exist for different lattice geometries, indicating that ENZ excitations are possible in a wide variety of systems. 

Standard macroscopic electrodynamics---and the associated concepts of refractive index and other bulk medium parameters---is based on a continuous effective-medium mean-field theory. This framework assumes that each atom interacts only with the average behavior of its surrounding atoms, thereby erasing spatial information about the position-dependent resonant dipole-dipole interactions between the atoms~\cite{Javanainen2014a,Javanainen17}. In contrast, here we have extracted the effective medium refractive index by directly analyzing the phase variation of the microscopic solutions.

To highlight how such approaches can substantially differ, we represent the lattice as a \textit{continuous} medium of the same thickness with continuous atomic polarization density $\bm{\mathcal{P}}(x)$, and consider the analogous standard optics problem~\cite{Jackson,BOR99}. The solution with the boundary conditions at the entrance and exit interfaces then yields $\bm{\mathcal{P}}(x)$ and light amplitudes in terms of complex $k’$, according to Eqs.~\eqref{eq:planelight} and~\eqref{eq:planepol}, including the reflected and transmitted light as a function of the medium thickness. Continuous-media electrodynamics is then consistent with solving the wavenumber from $k'^2/k^2-1=\alpha\rho/(1-\alpha\rho/3)$, as a function of the atomic polarizability $\alpha$ and density $\rho$~\cite{Javanainen2014a,Javanainen17}.

Although the continuous-media theory permits solutions where ${\rm Re}(k’)\rightarrow0$, this occurs only with significant losses ${\rm Im}(k’)>0$, and at densities so high that the continuous-medium approximation qualitatively breaks down~\cite{Javanainen2014a,JavanainenMFT}.  While losses can be suppressed by interferences  in multilevel transitions, this introduces additional complexity~\cite{McCutcheon24}.  At the same atom density as in Fig.~\ref{fig:fig2}, the continuous effective-medium mean-field theory predicts $k'/k\simeq 0.99+i0.13$ and vanishing transmission across the displayed frequency range,  dramatically deviating from the exact result.

If instead we take the numerically calculated exact reflection and transmission amplitudes of the granular medium and treat the atomic sample as continuous ``black box’’ medium governed by Eqs.~\eqref{eq:planelight} and~\eqref{eq:planepol}, we can solve for $k’$ from the resulting equations~\cite{BOR99,SmithEtAlPRB2002}. However, this approach still yields an incorrect refractive index, as shown in Fig.~\ref{Fig:fig5} in End Matter. Crucially, these discrepancies persist even as the number of layers increases. Although more advanced approaches~\cite{Alu2011} may yield a better matching, the deviations indicate a fundamental breakdown of common continuous-media approximations in capturing collective resonance effects near the ENZ regime.

So far, we have considered atoms at fixed positions. Numerically, we can study the effects of atomic position fluctuations in small lattices.  For a six-layer array, we find that the ENZ behavior remains robust under fluctuations. In each stochastic realization, we sample atomic positions fluctuating around their mean lattice site positions with a root-mean-square width of $0.075a$~\cite{Jenkins2012a,Ruostekoski2023}. This corresponds to moderate optical lattice depths of several hundred recoil energies, as employed in light transmission experiments of Ref.~\cite{Rui2020}. The exact optical response is then obtained by ensemble-averaging over individual realizations~\cite{Javanainen1999a,Lee16}.

At $\Delta = 0.05\gamma$, we find the effective wavenumber of the atomic polarization amplitudes $\left<{\cal P}_j\right>$ at the central sites in the fluctuating case. The resulting value, $|{\rm Re}(k')/k|\simeq0.17\pm0.01$, does not deviate appreciably from the fixed position case $|{\rm Re}(k')/k|\simeq0.19$. The dominant effect of position fluctuations is a smoothing of the transmission lineshape.   

We have shown how an effective ENZ response---with long-wavelength excitations and suppressed phase accumulation---can emerge from collective light scattering in a discrete atomic lattice, based on essentially exact microscopic simulations. These results provide a concrete, atomistic demonstration of how macroscopic optical behavior can arise from first-principles light-matter interactions, without relying on phenomenological effective-medium models. The framework presented here opens the door to designing and exploring quantum optics in a unique ENZ environment. 

One long-standing challenge in this context is achieving uniform optical driving across a 3D atomic ensemble. In free space, each atom experiences a different phase of the incident and scattered fields, while idealized cooperative models--- such as the Dicke superradiance~\cite{Dicke54} and related models undergoing quantum phase transitions~\cite{Walls1978,Carmichael1980,Ferioli2023,Agarwal2024,Ruostekoski25,Goncalves25}---typically assume the absence of any such spatial phase variation. An ENZ medium, with its nearly constant phase profile across the sample, could provide an ideal platform to realize such cooperative behavior. Additionally, the characteristic length scale $1/k$ of light-mediated dipole–dipole interactions in free space~\cite{Javanainen2014a,JavanainenMFT} necessitates experimentally challenging~\cite{Olmos13} far below subwavelength spacing to reach optically-established strongly correlated regimes~\cite{Olmos16,Williamson2020b,cidrim2020,Parmee2020,scarlatella2024,tecer2025} or topologically nontrivial phases~\cite{Perczel2017a,Bettles_topo}. In contrast, an ENZ medium could potentially extend the length scale to $1/k’$, substantially strengthening interactions. In quantum information applications, this can translate into enhanced light-mediated long-range entanglement generation between qubits~\cite{Biehs17,li19}.

Beyond these interaction effects, ENZ media may enable new approaches to precision measurement. Just as negative-index materials give rise to inverse Doppler shifts~\cite{Chen2011}, ENZ media may suppress Doppler shifts---one of the key limitations in high-precision spectroscopy. Additional exotic ENZ-induced emitter phenomena have been suggested~\cite{Lobet20}, and the rapidly developing field of time-varying photonics~\cite{Zhou20,Lozano23,Tirole23} offers further exciting avenues for ENZ-based quantum control.

\begin{acknowledgments}
We acknowledge financial support, in part, from Moonshot R\&D, JST JPMJMS2061 (L.R.) and EPSRC (Grant No.\ EP/S002952/1) (J.R.). The potential connection of the ENZ response and Doppler shift cancellation was suggested by W.D.\ Phillips. 
\end{acknowledgments}

\section{End Matter}
\appendix

\noindent\textit{Light propagation through array planes}:
The propagation of light can be efficiently calculated by exploiting the translational invariance of each atomic layer in Eq.~\eqref{eq:layer-layer-scattering}. Our approach follows a procedure similar to that introduced in Ref.~\cite{ruks25} for evaluating negative refraction for stacked atomic layers. To derive Eq.~\eqref{eq:layer-layer-scattering}, the excitation of each atom $\textbf{n}$ in Eq.~\eqref{eq:point-scattering-main} is expressed as a Bloch-wave sum within the corresponding layer $\ell$,  $\mathcal{P}^{(\textbf{n})} = \sum_{\textbf{q}_{\parallel}}\bm{\mathcal{P}}_{\ell}(\textbf{q}_{\parallel})e^{i(q_{\parallel y} y_{\textbf{n}} + q_{\parallel z} z_{\textbf{n}})}$. 
The interaction term in Eq.~\eqref{eq:point-scattering-main} then separates into intralayer and interlayer radiative couplings between phase-matched Bloch waves. To evaluate these terms, we expand the dipole radiation kernel ${\sf G}_{\nu\mu}({\bf r})$, given by
\begin{equation}
    {\sf G}_{\nu\mu}({\bf r}) = \left( {\partial\over\partial r_\nu}{\partial\over\partial r_\mu} -
\delta_{\nu\mu} \boldsymbol{\nabla}^2\right) {e^{ikr}\over4\pi r}
-\delta_{\nu\mu}\delta({\bf r}),
\label{eq:rad-kernel}
\end{equation}
as a sum of plane waves using the Weyl identity~\cite{Greffet_nanophotonics,Novotny_nanooptics,Belov05,belov06,Shahmoon}:
\begin{align}
{\sf G}_{\nu\mu}({\bf r}) 
&= \frac{i}{8\pi^2}  \int {d^2\textbf{q}_\parallel} \left( {\partial\over\partial r_\nu}{\partial\over\partial r_\mu} +
\delta_{\nu\mu} k^2\right)\frac{1}{k_{\perp}}{e^{i \textbf{q}_\parallel\cdot\textbf{r}_{\parallel}} e^{i k_\perp |x|}},
\label{eq:GDF2}
\end{align}
where $k_{\perp} = \sqrt{k^2 - q_{\parallel}^2}$, and
the integrand is recognized as the 2D Fourier transform of the radiation kernel on the array plane:
\begin{equation}
    {\sf G}({\bf r}) =\int \frac{d^2\textbf{q}_\parallel}{(2\pi)^2}e^{i\textbf{q}_{\parallel}\cdot\textbf{r}_{\parallel}} \tilde{\sf G}^{\parallel}(\textbf{q}_{\parallel}).
    \label{eq:fourier-transform}
\end{equation}
By substituting Eq.~\eqref{eq:GDF2} into the  interaction term in Eq.~\eqref{eq:point-scattering-main}, and evaluating the lattice sum over  in-plane atomic positions $j$ using the 2D Poisson summation formula, we obtain an integral over in-plane quasi-momenta.
This yields the interlayer interaction in Eq.~\eqref{eq:layer-layer-scattering}, with the layer propagator defined as
\begin{align}
    \mathsf{G}^{\mathrm{L}}_{\nu\mu}(\textbf{r},\qv_{\parallel}) &= \frac{1}{a^2}\sum_{j}\tilde{\sf G}^{\parallel}_{\nu\mu}(\textbf{k}_{\parallel} + \textbf{g}_{j})e^{i(\textbf{k}_{\parallel} + \textbf{g}_{j})\cdot \textbf{r}_{\parallel}} \nonumber\\&= \frac{i}{2a^2} \sum_j \frac{1}{k_\perp({\bf g}_j)}\left(- k_{j\nu}^{\text{B}}k_{j\mu}^{\text{B}}  +\delta_{\nu\mu} k^2 \right)e^{i\textbf{k}_{j}^{\mathrm{B}}\cdot \textbf{r}},
\end{align}
where each term in the sum corresponds to a projection onto the subspace orthogonal to the $j$th Bragg order wavevector $\textbf{k}_{j}^{\text{B}} = [\text{sgn}(x)k_{\perp}( \textbf{g}_{j}),\textbf{q}_{\parallel} + \textbf{g}_{j}]$, with $k_\perp({\bf g}_j) = { \sqrt{k^2-|\textbf{q}_{\parallel} + \textbf{g}_{j}|^2}}$. The 2D reciprocal lattice vectors $\mathbf{g}_{j}$ satisfy the periodicity condition $e^{i\mathbf{g}_{j}\cdot\mathbf{r}_{\parallel \ell}} = 1$.

The sum over Bragg orders in the layer propagator reflects the presence of both propagating and evanescent components of the dipole field. For subwavelength spacings, evanescent waves are contained in the higher-order terms,  contributing to near-field coupling between adjacent layers. These components are absent in 1D electrodynamics but are crucial in capturing the full microscopic scattering behavior in discrete lattices~\cite{Javanainen19}. The intralayer coupling in Eq.~\eqref{eq:point-scattering-main} is similarly evaluated in momentum space
\begin{align} \label{eq:transformX}
&\sum_{j \neq 0}e^{i\qv_{\parallel}\cdot \mathbf{r}_{\parallel j}}\mathsf{G}(\textbf{r}_{\parallel j})  =  \frac{1}{a^2}\sum_j \tilde{\mathsf{G}}^{\parallel}(\textbf{q}_{\parallel}+{\bf g}_j) -   {\mathsf{G}}(0),
\end{align}
where the formally divergent term is regularized~\cite{castin09,Antezza09b,Perczel2017a} by introducing a cutoff parameter $\eta$ in the Fourier transform,
\begin{equation}
\label{eq:regularization}
\tilde{\mathsf{G}}^{\parallel *}_{\nu \mu}(\textbf{q}_{\parallel}) = \int \frac{dq_{\perp}}{2\pi k^2}\frac{k^2\delta_{\nu\mu} - q_{\nu}q_{\mu}}{k^2 - (q_{\parallel}^2 + q_{\perp}^2) + i\epsilon}e^{-q^2\eta^2/4}e^{iq_{\perp}x},
\end{equation}
for $\epsilon>0$ and taking the limit $\eta \to 0$ in Eq.~\eqref{eq:transformX}. 
This completes the derivation of  Eq.~\eqref{eq:layer-layer-scattering}.\\

\noindent\textit{Collective resonance bands}:
Collective resonance linewidths and line shifts may more generally be obtained by expressing Eq.~\eqref{eq:point-scattering-main} in a matrix form~\cite{Ruostekoski2023}:
\begin{equation}
    \dot{\mathbf{b}} = i\left(\bm{\mathcal{H}} + \delta\bm{\mathcal{H}}\right)\mathbf{b} + \mathbf{f},
\end{equation}
where $\mathbf{b}_{3\textbf{n} - 1 + \nu} = \mathcal{P}_{\nu}^{(\textbf{n})},$ $\mathbf{f}_{3\textbf{n} - 1 + \nu} = i\hat{\textbf{e}}_{\nu}^{*} \cdot \bm{\mathcal{R}}^{+}(\textbf{r}_{\textbf{n}}),$ the diagonal elements of $\bm{\mathcal{H}}$ are $i\gamma$, and the diagonal matrix $\delta\bm{\mathcal{H}}$ contains $\Delta.$ The nondiagonal elements coupling atomic dipoles read 
\begin{equation}
    \mathcal{H}_{3\textbf{n} - 1 + \nu, 3\textbf{n}' - 1 + \mu} = \xi\mathsf{G}_{\nu\mu}(\textbf{r}_{\textbf{n}} - \textbf{r}_{\textbf{n}'}), \quad \textbf{n} \neq \textbf{n}'.
\end{equation}
The eigenvalues $\lambda_{j} = \delta_j + i\upsilon_j$ of $\bm{\mathcal{H}}$ contain the line shift $\delta_{j}$, from single-atom resonance, and the linewidth $\upsilon_{j}$ of collective atomic excitations, with eigenvectors describing the dipole amplitudes~\cite{Ruostekoski2023}. 

For infinitely many layers ($N_{x} = \infty$), a Bloch-wave ansatz for the dipole amplitudes,  with quasimomentum $\textbf{q}$, gives rise to eigenmodes in an analogous $3\times 3$ linear equation,
\begin{equation}
    \dot{\mathbf{b}}(\textbf{q}) = i\left[\bm{\mathcal{H}}(\textbf{q}) + \delta\bm{\mathcal{H}}\right]\mathbf{b}(\textbf{q}) + \textbf{f}(\textbf{q}).
\end{equation}
with $\mathbf{b}_{\nu}(\textbf{q}) = \bm{\mathcal{P}}^{(\textbf{n})}_{\nu}e^{-i\textbf{q}\cdot\textbf{r}_{\textbf{n}}}$ and $\mathbf{f}_{\nu}(\textbf{q}) = \bm{\mathcal{R}}_{\nu}(\textbf{r}_{\textbf{n}})e^{-i\textbf{q}\cdot\textbf{r}_{\textbf{n}}}.$ $\bm{\mathcal{H}}(\textbf{q})$ (and $\delta\bm{\mathcal{H}}$) is now a $3 \times 3$ matrix,
\begin{equation}
{\mathcal{H}}_{\nu\mu}(\textbf{q}) = \xi\sum_{\textbf{n} \neq \textbf{0}}e^{i\textbf{q}\cdot\textbf{r}_{\textbf{n}}}\mathsf{G}_{\nu\mu}(\textbf{r}_{\textbf{n}}) - i\gamma\delta_{\nu\mu},
\end{equation}
which is evaluated~\cite{castin09,Antezza09b} similarly to Eq.~\eqref{eq:transformX}. Its eigenvalues, $\delta^{(j)}(\textbf{q})$, are purely real owing to the absence of radiating boundaries, giving the three continuous bands of line shifts presented in Fig.~\ref{fig:fig4}.\\

\begin{figure}[h!]
    \centering \includegraphics[width=\linewidth]{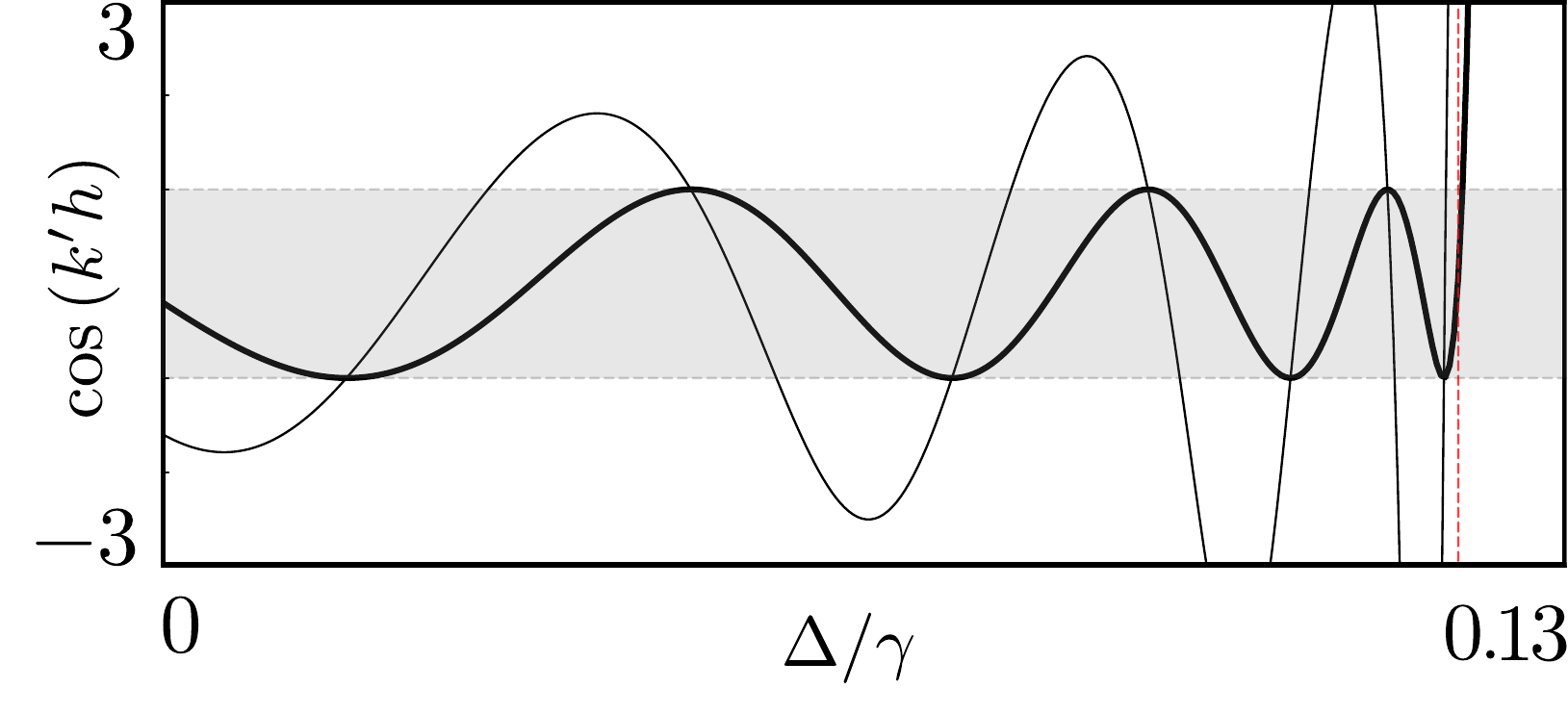}
    \caption{
(Real) phase parameter 
calculated using the wavenumber $k'$ (thick line) from the exact simulations [Fig.\ref{fig:fig2}(a)] and a fit based on the reflection and transmission amplitudes of the exact simulations (thin line). The region $\left|\cos\left(k'h\right)\right| \leq 1$ is shaded and the vertical red dashed line denotes $\Delta \simeq 0.12\gamma$, where $k' \simeq 0$ [Fig.~\ref{fig:fig2}(b)].}
    \label{Fig:fig5}
\end{figure}

\noindent\textit{Approximate medium propagation}: We showed that the effective mean-field theory of standard textbook continuous-media optics fails to capture the optical response of the atomic array. We can alternatively employ the reflection and transmission amplitudes of the exact numerical simulations of the atomic array and match these with the continuous-medium ansatz [Eqs.~\eqref{eq:planelight} and~\eqref{eq:planepol}] to extract $k’$. Here, we find that $\cos \left(k' h\right)$ is real-valued and we use it to quantify phase accumulation across the medium [when $\cos \left(k' h\right)<1$] or attenuation [when $\cos \left(k' h\right)>1$]. Even this approach leads to an incorrect effective refractive index---including an erroneous prediction of ${\rm Im}(k’)$---as displayed in Fig.~\ref{Fig:fig5}. \\

\begin{figure}[h!]
    \centering
    \includegraphics[width=\linewidth]{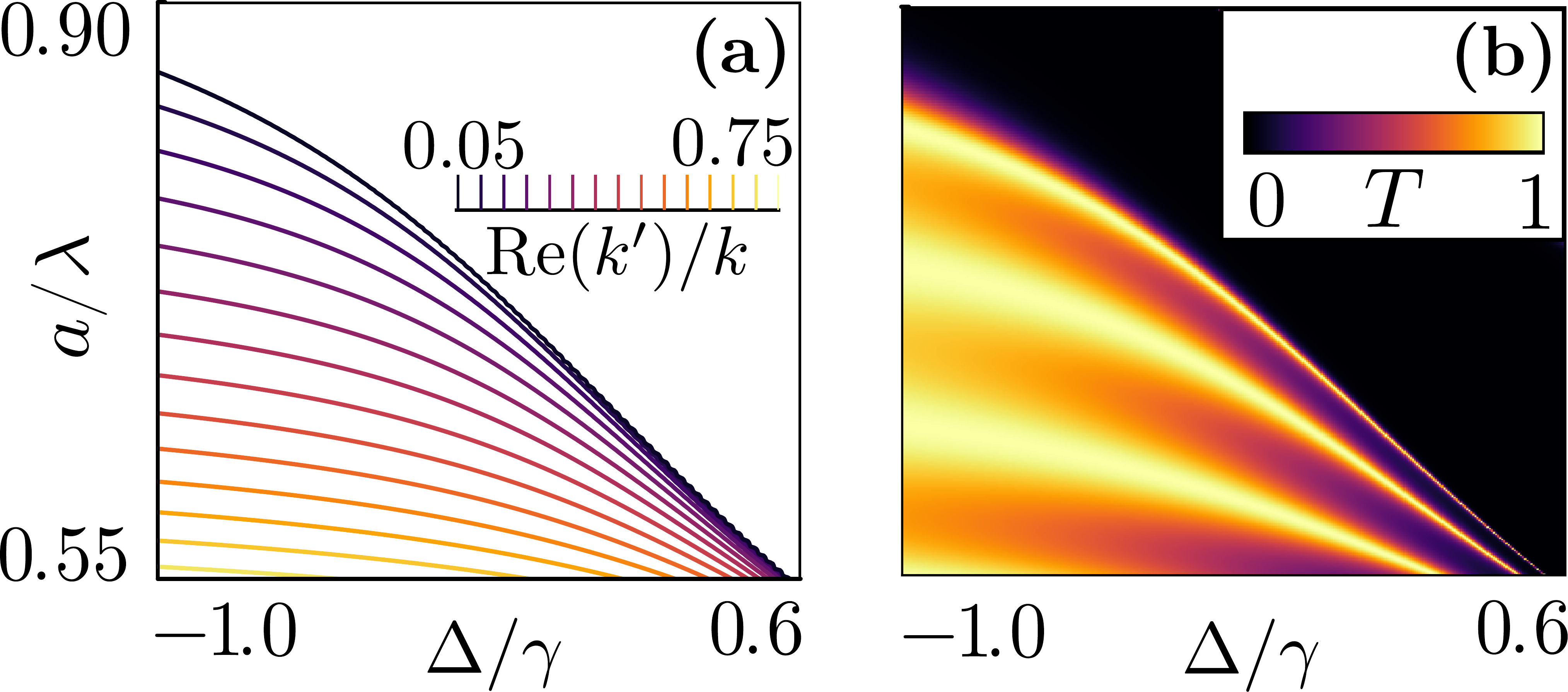}
    \caption{(a) Real part of the medium wavenumber $k'$ and (b) power transmission $T$ for five layers, as detuning $\Delta$ and lattice spacing $a$ are varied. Within the transmission band we have $\mathrm{Im}(k') \simeq 0$.}
    \label{fig:fig6}
\end{figure}

\noindent\textit{ENZ regime in varying lattices:} Although we focused on the lattice with $a = 0.66\lambda,$ $N_{x} = 25$, an ENZ response is present over a range of lattice spacings and layer numbers. Figure~\ref{fig:fig6}(a) shows that for $N_{x} = 5$, the medium wavenumber approaches zero for a range of values $a$, with only quantitative changes in the ENZ regime. The edge of the transmission band [Fig.~\ref{fig:fig6}(b)] is similarly found to accompany the smallest wavenumbers.
While the number of Fano resonances increases linearly with layer number, the wavenumbers observed in Fig.~\ref{fig:fig6}(a) remain essentially unchanged. We note that, as suggested by Fig. \ref{fig:fig4}, a second transmission band is generally found as detuning is increased further through the reflecting region. This band instead comprises the most rapidly varying collective excitations with $k'$ in the vicinity of $\pi/a,$ and is not relevant to the present study.

\begin{thebibliography}{95}%
\makeatletter
\providecommand \@ifxundefined [1]{%
 \@ifx{#1\undefined}
}%
\providecommand \@ifnum [1]{%
 \ifnum #1\expandafter \@firstoftwo
 \else \expandafter \@secondoftwo
 \fi
}%
\providecommand \@ifx [1]{%
 \ifx #1\expandafter \@firstoftwo
 \else \expandafter \@secondoftwo
 \fi
}%
\providecommand \natexlab [1]{#1}%
\providecommand \enquote  [1]{``#1''}%
\providecommand \bibnamefont  [1]{#1}%
\providecommand \bibfnamefont [1]{#1}%
\providecommand \citenamefont [1]{#1}%
\providecommand \href@noop [0]{\@secondoftwo}%
\providecommand \href [0]{\begingroup \@sanitize@url \@href}%
\providecommand \@href[1]{\@@startlink{#1}\@@href}%
\providecommand \@@href[1]{\endgroup#1\@@endlink}%
\providecommand \@sanitize@url [0]{\catcode `\\12\catcode `\$12\catcode
  `\&12\catcode `\#12\catcode `\^12\catcode `\_12\catcode `\%12\relax}%
\providecommand \@@startlink[1]{}%
\providecommand \@@endlink[0]{}%
\providecommand \url  [0]{\begingroup\@sanitize@url \@url }%
\providecommand \@url [1]{\endgroup\@href {#1}{\urlprefix }}%
\providecommand \urlprefix  [0]{URL }%
\providecommand \Eprint [0]{\href }%
\providecommand \doibase [0]{https://doi.org/}%
\providecommand \selectlanguage [0]{\@gobble}%
\providecommand \bibinfo  [0]{\@secondoftwo}%
\providecommand \bibfield  [0]{\@secondoftwo}%
\providecommand \translation [1]{[#1]}%
\providecommand \BibitemOpen [0]{}%
\providecommand \bibitemStop [0]{}%
\providecommand \bibitemNoStop [0]{.\EOS\space}%
\providecommand \EOS [0]{\spacefactor3000\relax}%
\providecommand \BibitemShut  [1]{\csname bibitem#1\endcsname}%
\let\auto@bib@innerbib\@empty
\bibitem [{\citenamefont {Jackson}(1999)}]{Jackson}%
  \BibitemOpen
  \bibfield  {author} {\bibinfo {author} {\bibfnamefont {J.~D.}\ \bibnamefont
  {Jackson}},\ }\href@noop {} {\emph {\bibinfo {title} {Classical
  Electrodynamics}}},\ \bibinfo {edition} {3rd}\ ed.\ (\bibinfo  {publisher}
  {Wiley, New York},\ \bibinfo {year} {1999})\BibitemShut {NoStop}%
\bibitem [{\citenamefont {Born}\ and\ \citenamefont {Wolf}(1999)}]{BOR99}%
  \BibitemOpen
  \bibfield  {author} {\bibinfo {author} {\bibfnamefont {M.}~\bibnamefont
  {Born}}\ and\ \bibinfo {author} {\bibfnamefont {E.}~\bibnamefont {Wolf}},\
  }\href@noop {} {\emph {\bibinfo {title} {Principles of Optics}}},\ \bibinfo
  {edition} {7th}\ ed.\ (\bibinfo  {publisher} {Cambridge University Press,
  Cambridge, UK},\ \bibinfo {year} {1999})\BibitemShut {NoStop}%
\bibitem [{\citenamefont {Javanainen}\ \emph {et~al.}(2014)\citenamefont
  {Javanainen}, \citenamefont {Ruostekoski}, \citenamefont {Li},\ and\
  \citenamefont {Yoo}}]{Javanainen2014a}%
  \BibitemOpen
  \bibfield  {author} {\bibinfo {author} {\bibfnamefont {J.}~\bibnamefont
  {Javanainen}}, \bibinfo {author} {\bibfnamefont {J.}~\bibnamefont
  {Ruostekoski}}, \bibinfo {author} {\bibfnamefont {Y.}~\bibnamefont {Li}},\
  and\ \bibinfo {author} {\bibfnamefont {S.-M.}\ \bibnamefont {Yoo}},\
  }\bibfield  {title} {\bibinfo {title} {Shifts of a resonance line in a dense
  atomic sample},\ }\href {https://doi.org/10.1103/PhysRevLett.112.113603}
  {\bibfield  {journal} {\bibinfo  {journal} {Phys. Rev. Lett.}\ }\textbf
  {\bibinfo {volume} {112}},\ \bibinfo {pages} {113603} (\bibinfo {year}
  {2014})}\BibitemShut {NoStop}%
\bibitem [{\citenamefont {Javanainen}\ and\ \citenamefont
  {Ruostekoski}(2016)}]{JavanainenMFT}%
  \BibitemOpen
  \bibfield  {author} {\bibinfo {author} {\bibfnamefont {J.}~\bibnamefont
  {Javanainen}}\ and\ \bibinfo {author} {\bibfnamefont {J.}~\bibnamefont
  {Ruostekoski}},\ }\bibfield  {title} {\bibinfo {title} {Light propagation
  beyond the mean-field theory of standard optics},\ }\href
  {https://doi.org/10.1364/OE.24.000993} {\bibfield  {journal} {\bibinfo
  {journal} {Opt. Express}\ }\textbf {\bibinfo {volume} {24}},\ \bibinfo
  {pages} {993} (\bibinfo {year} {2016})}\BibitemShut {NoStop}%
\bibitem [{\citenamefont {Jenkins}\ \emph {et~al.}(2016)\citenamefont
  {Jenkins}, \citenamefont {Ruostekoski}, \citenamefont {Javanainen},
  \citenamefont {Bourgain}, \citenamefont {Jennewein}, \citenamefont
  {Sortais},\ and\ \citenamefont {Browaeys}}]{Jenkins_thermshift}%
  \BibitemOpen
  \bibfield  {author} {\bibinfo {author} {\bibfnamefont {S.~D.}\ \bibnamefont
  {Jenkins}}, \bibinfo {author} {\bibfnamefont {J.}~\bibnamefont
  {Ruostekoski}}, \bibinfo {author} {\bibfnamefont {J.}~\bibnamefont
  {Javanainen}}, \bibinfo {author} {\bibfnamefont {R.}~\bibnamefont
  {Bourgain}}, \bibinfo {author} {\bibfnamefont {S.}~\bibnamefont {Jennewein}},
  \bibinfo {author} {\bibfnamefont {Y.~R.~P.}\ \bibnamefont {Sortais}},\ and\
  \bibinfo {author} {\bibfnamefont {A.}~\bibnamefont {Browaeys}},\ }\bibfield
  {title} {\bibinfo {title} {Optical resonance shifts in the fluorescence of
  thermal and cold atomic gases},\ }\href
  {https://doi.org/10.1103/PhysRevLett.116.183601} {\bibfield  {journal}
  {\bibinfo  {journal} {Phys. Rev. Lett.}\ }\textbf {\bibinfo {volume} {116}},\
  \bibinfo {pages} {183601} (\bibinfo {year} {2016})}\BibitemShut {NoStop}%
\bibitem [{\citenamefont {Andreoli}\ \emph {et~al.}(2021)\citenamefont
  {Andreoli}, \citenamefont {Gullans}, \citenamefont {High}, \citenamefont
  {Browaeys},\ and\ \citenamefont {Chang}}]{Andreoli21}%
  \BibitemOpen
  \bibfield  {author} {\bibinfo {author} {\bibfnamefont {F.}~\bibnamefont
  {Andreoli}}, \bibinfo {author} {\bibfnamefont {M.~J.}\ \bibnamefont
  {Gullans}}, \bibinfo {author} {\bibfnamefont {A.~A.}\ \bibnamefont {High}},
  \bibinfo {author} {\bibfnamefont {A.}~\bibnamefont {Browaeys}},\ and\
  \bibinfo {author} {\bibfnamefont {D.~E.}\ \bibnamefont {Chang}},\ }\bibfield
  {title} {\bibinfo {title} {Maximum refractive index of an atomic medium},\
  }\href {https://doi.org/10.1103/PhysRevX.11.011026} {\bibfield  {journal}
  {\bibinfo  {journal} {Phys. Rev. X}\ }\textbf {\bibinfo {volume} {11}},\
  \bibinfo {pages} {011026} (\bibinfo {year} {2021})}\BibitemShut {NoStop}%
\bibitem [{\citenamefont {Liberal}\ and\ \citenamefont
  {Engheta}(2017{\natexlab{a}})}]{Liberal17}%
  \BibitemOpen
  \bibfield  {author} {\bibinfo {author} {\bibfnamefont {I.}~\bibnamefont
  {Liberal}}\ and\ \bibinfo {author} {\bibfnamefont {N.}~\bibnamefont
  {Engheta}},\ }\bibfield  {title} {\bibinfo {title} {Near-zero refractive
  index photonics},\ }\href {https://doi.org/10.1038/nphoton.2017.13}
  {\bibfield  {journal} {\bibinfo  {journal} {Nature Photonics}\ }\textbf
  {\bibinfo {volume} {11}},\ \bibinfo {pages} {149} (\bibinfo {year}
  {2017}{\natexlab{a}})}\BibitemShut {NoStop}%
\bibitem [{\citenamefont {Lobet}\ \emph {et~al.}(2023)\citenamefont {Lobet},
  \citenamefont {Kinsey}, \citenamefont {Liberal}, \citenamefont {Caglayan},
  \citenamefont {Huidobro}, \citenamefont {Galiffi}, \citenamefont
  {Mejía-Salazar}, \citenamefont {Palermo}, \citenamefont {Jacob},\ and\
  \citenamefont {Maccaferri}}]{Lobet23}%
  \BibitemOpen
  \bibfield  {author} {\bibinfo {author} {\bibfnamefont {M.}~\bibnamefont
  {Lobet}}, \bibinfo {author} {\bibfnamefont {N.}~\bibnamefont {Kinsey}},
  \bibinfo {author} {\bibfnamefont {I.}~\bibnamefont {Liberal}}, \bibinfo
  {author} {\bibfnamefont {H.}~\bibnamefont {Caglayan}}, \bibinfo {author}
  {\bibfnamefont {P.~A.}\ \bibnamefont {Huidobro}}, \bibinfo {author}
  {\bibfnamefont {E.}~\bibnamefont {Galiffi}}, \bibinfo {author} {\bibfnamefont
  {J.~R.}\ \bibnamefont {Mejía-Salazar}}, \bibinfo {author} {\bibfnamefont
  {G.}~\bibnamefont {Palermo}}, \bibinfo {author} {\bibfnamefont
  {Z.}~\bibnamefont {Jacob}},\ and\ \bibinfo {author} {\bibfnamefont
  {N.}~\bibnamefont {Maccaferri}},\ }\bibfield  {title} {\bibinfo {title} {New
  horizons in near-zero refractive index photonics and hyperbolic
  metamaterials},\ }\href {https://doi.org/10.1021/acsphotonics.3c00747}
  {\bibfield  {journal} {\bibinfo  {journal} {ACS Photonics}\ }\textbf
  {\bibinfo {volume} {10}},\ \bibinfo {pages} {3805} (\bibinfo {year}
  {2023})}\BibitemShut {NoStop}%
\bibitem [{\citenamefont {Xie}\ \emph {et~al.}(2025)\citenamefont {Xie},
  \citenamefont {Wang},\ and\ \citenamefont {Kivshar}}]{Xie2025}%
  \BibitemOpen
  \bibfield  {author} {\bibinfo {author} {\bibfnamefont {P.}~\bibnamefont
  {Xie}}, \bibinfo {author} {\bibfnamefont {W.}~\bibnamefont {Wang}},\ and\
  \bibinfo {author} {\bibfnamefont {Y.}~\bibnamefont {Kivshar}},\ }\bibfield
  {title} {\bibinfo {title} {Resonant light--matter interaction with
  epsilon-near-zero photonic structures},\ }\href
  {https://pubs.aip.org/aip/apr/article/12/2/021307/3344080} {\bibfield
  {journal} {\bibinfo  {journal} {Applied Physics Reviews}\ }\textbf {\bibinfo
  {volume} {12}} (\bibinfo {year} {2025})}\BibitemShut {NoStop}%
\bibitem [{\citenamefont {Enoch}\ \emph {et~al.}(2002)\citenamefont {Enoch},
  \citenamefont {Tayeb}, \citenamefont {Sabouroux}, \citenamefont {Gu\'erin},\
  and\ \citenamefont {Vincent}}]{Enoch02}%
  \BibitemOpen
  \bibfield  {author} {\bibinfo {author} {\bibfnamefont {S.}~\bibnamefont
  {Enoch}}, \bibinfo {author} {\bibfnamefont {G.}~\bibnamefont {Tayeb}},
  \bibinfo {author} {\bibfnamefont {P.}~\bibnamefont {Sabouroux}}, \bibinfo
  {author} {\bibfnamefont {N.}~\bibnamefont {Gu\'erin}},\ and\ \bibinfo
  {author} {\bibfnamefont {P.}~\bibnamefont {Vincent}},\ }\bibfield  {title}
  {\bibinfo {title} {A metamaterial for directive emission},\ }\href
  {https://doi.org/10.1103/PhysRevLett.89.213902} {\bibfield  {journal}
  {\bibinfo  {journal} {Phys. Rev. Lett.}\ }\textbf {\bibinfo {volume} {89}},\
  \bibinfo {pages} {213902} (\bibinfo {year} {2002})}\BibitemShut {NoStop}%
\bibitem [{\citenamefont {Al\`u}\ \emph {et~al.}(2007)\citenamefont {Al\`u},
  \citenamefont {Silveirinha}, \citenamefont {Salandrino},\ and\ \citenamefont
  {Engheta}}]{Alu07}%
  \BibitemOpen
  \bibfield  {author} {\bibinfo {author} {\bibfnamefont {A.}~\bibnamefont
  {Al\`u}}, \bibinfo {author} {\bibfnamefont {M.~G.}\ \bibnamefont
  {Silveirinha}}, \bibinfo {author} {\bibfnamefont {A.}~\bibnamefont
  {Salandrino}},\ and\ \bibinfo {author} {\bibfnamefont {N.}~\bibnamefont
  {Engheta}},\ }\bibfield  {title} {\bibinfo {title} {Epsilon-near-zero
  metamaterials and electromagnetic sources: Tailoring the radiation phase
  pattern},\ }\href {https://doi.org/10.1103/PhysRevB.75.155410} {\bibfield
  {journal} {\bibinfo  {journal} {Phys. Rev. B}\ }\textbf {\bibinfo {volume}
  {75}},\ \bibinfo {pages} {155410} (\bibinfo {year} {2007})}\BibitemShut
  {NoStop}%
\bibitem [{\citenamefont {Pacheco-Peña}\ \emph {et~al.}(2014)\citenamefont
  {Pacheco-Peña}, \citenamefont {Torres}, \citenamefont {Orazbayev},
  \citenamefont {Beruete}, \citenamefont {Navarro-Cía}, \citenamefont
  {Sorolla},\ and\ \citenamefont {Engheta}}]{Pacheco14}%
  \BibitemOpen
  \bibfield  {author} {\bibinfo {author} {\bibfnamefont {V.}~\bibnamefont
  {Pacheco-Peña}}, \bibinfo {author} {\bibfnamefont {V.}~\bibnamefont
  {Torres}}, \bibinfo {author} {\bibfnamefont {B.}~\bibnamefont {Orazbayev}},
  \bibinfo {author} {\bibfnamefont {M.}~\bibnamefont {Beruete}}, \bibinfo
  {author} {\bibfnamefont {M.}~\bibnamefont {Navarro-Cía}}, \bibinfo {author}
  {\bibfnamefont {M.}~\bibnamefont {Sorolla}},\ and\ \bibinfo {author}
  {\bibfnamefont {N.}~\bibnamefont {Engheta}},\ }\bibfield  {title} {\bibinfo
  {title} {Mechanical 144ghz beam steering with all-metallic epsilon-near-zero
  lens antenna},\ }\href {https://doi.org/10.1063/1.4903865} {\bibfield
  {journal} {\bibinfo  {journal} {Applied Physics Letters}\ }\textbf {\bibinfo
  {volume} {105}},\ \bibinfo {pages} {243503} (\bibinfo {year}
  {2014})}\BibitemShut {NoStop}%
\bibitem [{\citenamefont {Soric}\ and\ \citenamefont {Alù}(2015)}]{Soric15}%
  \BibitemOpen
  \bibfield  {author} {\bibinfo {author} {\bibfnamefont {J.~C.}\ \bibnamefont
  {Soric}}\ and\ \bibinfo {author} {\bibfnamefont {A.}~\bibnamefont {Alù}},\
  }\bibfield  {title} {\bibinfo {title} {Longitudinally independent matching
  and arbitrary wave patterning using $\varepsilon$ -near-zero channels},\
  }\href {https://doi.org/10.1109/TMTT.2015.2479589} {\bibfield  {journal}
  {\bibinfo  {journal} {IEEE Transactions on Microwave Theory and Techniques}\
  }\textbf {\bibinfo {volume} {63}},\ \bibinfo {pages} {3558} (\bibinfo {year}
  {2015})}\BibitemShut {NoStop}%
\bibitem [{\citenamefont {Silveirinha}\ and\ \citenamefont
  {Engheta}(2006)}]{Silveirinha06}%
  \BibitemOpen
  \bibfield  {author} {\bibinfo {author} {\bibfnamefont {M.}~\bibnamefont
  {Silveirinha}}\ and\ \bibinfo {author} {\bibfnamefont {N.}~\bibnamefont
  {Engheta}},\ }\bibfield  {title} {\bibinfo {title} {Tunneling of
  electromagnetic energy through subwavelength channels and bends using
  $\ensuremath{\epsilon}$-near-zero materials},\ }\href
  {https://doi.org/10.1103/PhysRevLett.97.157403} {\bibfield  {journal}
  {\bibinfo  {journal} {Phys. Rev. Lett.}\ }\textbf {\bibinfo {volume} {97}},\
  \bibinfo {pages} {157403} (\bibinfo {year} {2006})}\BibitemShut {NoStop}%
\bibitem [{\citenamefont {Edwards}\ \emph {et~al.}(2008)\citenamefont
  {Edwards}, \citenamefont {Al\`u}, \citenamefont {Young}, \citenamefont
  {Silveirinha},\ and\ \citenamefont {Engheta}}]{Edwards08}%
  \BibitemOpen
  \bibfield  {author} {\bibinfo {author} {\bibfnamefont {B.}~\bibnamefont
  {Edwards}}, \bibinfo {author} {\bibfnamefont {A.}~\bibnamefont {Al\`u}},
  \bibinfo {author} {\bibfnamefont {M.~E.}\ \bibnamefont {Young}}, \bibinfo
  {author} {\bibfnamefont {M.}~\bibnamefont {Silveirinha}},\ and\ \bibinfo
  {author} {\bibfnamefont {N.}~\bibnamefont {Engheta}},\ }\bibfield  {title}
  {\bibinfo {title} {Experimental verification of epsilon-near-zero
  metamaterial coupling and energy squeezing using a microwave waveguide},\
  }\href {https://doi.org/10.1103/PhysRevLett.100.033903} {\bibfield  {journal}
  {\bibinfo  {journal} {Phys. Rev. Lett.}\ }\textbf {\bibinfo {volume} {100}},\
  \bibinfo {pages} {033903} (\bibinfo {year} {2008})}\BibitemShut {NoStop}%
\bibitem [{\citenamefont {Adams}\ \emph {et~al.}(2011)\citenamefont {Adams},
  \citenamefont {Inampudi}, \citenamefont {Ribaudo}, \citenamefont {Slocum},
  \citenamefont {Vangala}, \citenamefont {Kuhta}, \citenamefont {Goodhue},
  \citenamefont {Podolskiy},\ and\ \citenamefont {Wasserman}}]{Adams11}%
  \BibitemOpen
  \bibfield  {author} {\bibinfo {author} {\bibfnamefont {D.~C.}\ \bibnamefont
  {Adams}}, \bibinfo {author} {\bibfnamefont {S.}~\bibnamefont {Inampudi}},
  \bibinfo {author} {\bibfnamefont {T.}~\bibnamefont {Ribaudo}}, \bibinfo
  {author} {\bibfnamefont {D.}~\bibnamefont {Slocum}}, \bibinfo {author}
  {\bibfnamefont {S.}~\bibnamefont {Vangala}}, \bibinfo {author} {\bibfnamefont
  {N.~A.}\ \bibnamefont {Kuhta}}, \bibinfo {author} {\bibfnamefont {W.~D.}\
  \bibnamefont {Goodhue}}, \bibinfo {author} {\bibfnamefont {V.~A.}\
  \bibnamefont {Podolskiy}},\ and\ \bibinfo {author} {\bibfnamefont
  {D.}~\bibnamefont {Wasserman}},\ }\bibfield  {title} {\bibinfo {title}
  {Funneling light through a subwavelength aperture with epsilon-near-zero
  materials},\ }\href {https://doi.org/10.1103/PhysRevLett.107.133901}
  {\bibfield  {journal} {\bibinfo  {journal} {Phys. Rev. Lett.}\ }\textbf
  {\bibinfo {volume} {107}},\ \bibinfo {pages} {133901} (\bibinfo {year}
  {2011})}\BibitemShut {NoStop}%
\bibitem [{\citenamefont {Argyropoulos}\ \emph {et~al.}(2012)\citenamefont
  {Argyropoulos}, \citenamefont {Chen}, \citenamefont {D'Aguanno},
  \citenamefont {Engheta},\ and\ \citenamefont {Al\`u}}]{Argyropoulos12}%
  \BibitemOpen
  \bibfield  {author} {\bibinfo {author} {\bibfnamefont {C.}~\bibnamefont
  {Argyropoulos}}, \bibinfo {author} {\bibfnamefont {P.-Y.}\ \bibnamefont
  {Chen}}, \bibinfo {author} {\bibfnamefont {G.}~\bibnamefont {D'Aguanno}},
  \bibinfo {author} {\bibfnamefont {N.}~\bibnamefont {Engheta}},\ and\ \bibinfo
  {author} {\bibfnamefont {A.}~\bibnamefont {Al\`u}},\ }\bibfield  {title}
  {\bibinfo {title} {Boosting optical nonlinearities in
  $\ensuremath{\epsilon}$-near-zero plasmonic channels},\ }\href
  {https://doi.org/10.1103/PhysRevB.85.045129} {\bibfield  {journal} {\bibinfo
  {journal} {Phys. Rev. B}\ }\textbf {\bibinfo {volume} {85}},\ \bibinfo
  {pages} {045129} (\bibinfo {year} {2012})}\BibitemShut {NoStop}%
\bibitem [{\citenamefont {Alam}\ \emph {et~al.}(2016)\citenamefont {Alam},
  \citenamefont {Leon},\ and\ \citenamefont {Boyd}}]{Zahirul16}%
  \BibitemOpen
  \bibfield  {author} {\bibinfo {author} {\bibfnamefont {M.~Z.}\ \bibnamefont
  {Alam}}, \bibinfo {author} {\bibfnamefont {I.~D.}\ \bibnamefont {Leon}},\
  and\ \bibinfo {author} {\bibfnamefont {R.~W.}\ \bibnamefont {Boyd}},\
  }\bibfield  {title} {\bibinfo {title} {Large optical nonlinearity of indium
  tin oxide in its epsilon-near-zero region},\ }\href
  {https://doi.org/10.1126/science.aae0330} {\bibfield  {journal} {\bibinfo
  {journal} {Science}\ }\textbf {\bibinfo {volume} {352}},\ \bibinfo {pages}
  {795} (\bibinfo {year} {2016})}\BibitemShut {NoStop}%
\bibitem [{\citenamefont {Reshef}\ \emph {et~al.}(2019)\citenamefont {Reshef},
  \citenamefont {De~Leon}, \citenamefont {Alam},\ and\ \citenamefont
  {Boyd}}]{Reshef19}%
  \BibitemOpen
  \bibfield  {author} {\bibinfo {author} {\bibfnamefont {O.}~\bibnamefont
  {Reshef}}, \bibinfo {author} {\bibfnamefont {I.}~\bibnamefont {De~Leon}},
  \bibinfo {author} {\bibfnamefont {M.~Z.}\ \bibnamefont {Alam}},\ and\
  \bibinfo {author} {\bibfnamefont {R.~W.}\ \bibnamefont {Boyd}},\ }\bibfield
  {title} {\bibinfo {title} {Nonlinear optical effects in epsilon-near-zero
  media},\ }\href {https://doi.org/10.1038/s41578-019-0120-5} {\bibfield
  {journal} {\bibinfo  {journal} {Nature Reviews Materials}\ }\textbf {\bibinfo
  {volume} {4}},\ \bibinfo {pages} {535} (\bibinfo {year} {2019})}\BibitemShut
  {NoStop}%
\bibitem [{\citenamefont {Yang}\ \emph {et~al.}(2019)\citenamefont {Yang},
  \citenamefont {Lu}, \citenamefont {Manjavacas}, \citenamefont {Luk},
  \citenamefont {Liu}, \citenamefont {Kelley}, \citenamefont {Maria},
  \citenamefont {Runnerstrom}, \citenamefont {Sinclair}, \citenamefont
  {Ghimire},\ and\ \citenamefont {Brener}}]{Yang19}%
  \BibitemOpen
  \bibfield  {author} {\bibinfo {author} {\bibfnamefont {Y.}~\bibnamefont
  {Yang}}, \bibinfo {author} {\bibfnamefont {J.}~\bibnamefont {Lu}}, \bibinfo
  {author} {\bibfnamefont {A.}~\bibnamefont {Manjavacas}}, \bibinfo {author}
  {\bibfnamefont {T.~S.}\ \bibnamefont {Luk}}, \bibinfo {author} {\bibfnamefont
  {H.}~\bibnamefont {Liu}}, \bibinfo {author} {\bibfnamefont {K.}~\bibnamefont
  {Kelley}}, \bibinfo {author} {\bibfnamefont {J.-P.}\ \bibnamefont {Maria}},
  \bibinfo {author} {\bibfnamefont {E.~L.}\ \bibnamefont {Runnerstrom}},
  \bibinfo {author} {\bibfnamefont {M.~B.}\ \bibnamefont {Sinclair}}, \bibinfo
  {author} {\bibfnamefont {S.}~\bibnamefont {Ghimire}},\ and\ \bibinfo {author}
  {\bibfnamefont {I.}~\bibnamefont {Brener}},\ }\bibfield  {title} {\bibinfo
  {title} {High-harmonic generation from an epsilon-near-zero material},\
  }\href {https://doi.org/10.1038/s41567-019-0584-7} {\bibfield  {journal}
  {\bibinfo  {journal} {Nature Physics}\ }\textbf {\bibinfo {volume} {15}},\
  \bibinfo {pages} {1022} (\bibinfo {year} {2019})}\BibitemShut {NoStop}%
\bibitem [{\citenamefont {Liberal}\ and\ \citenamefont
  {Engheta}(2017{\natexlab{b}})}]{Liberal17c}%
  \BibitemOpen
  \bibfield  {author} {\bibinfo {author} {\bibfnamefont {I.}~\bibnamefont
  {Liberal}}\ and\ \bibinfo {author} {\bibfnamefont {N.}~\bibnamefont
  {Engheta}},\ }\bibfield  {title} {\bibinfo {title} {Zero-index structures as
  an alternative platform for quantum optics},\ }\href
  {https://doi.org/10.1073/pnas.1611924114} {\bibfield  {journal} {\bibinfo
  {journal} {Proceedings of the National Academy of Sciences}\ }\textbf
  {\bibinfo {volume} {114}},\ \bibinfo {pages} {822} (\bibinfo {year}
  {2017}{\natexlab{b}})}\BibitemShut {NoStop}%
\bibitem [{\citenamefont {Campione}\ \emph {et~al.}(2015)\citenamefont
  {Campione}, \citenamefont {Liu}, \citenamefont {Benz}, \citenamefont {Klem},
  \citenamefont {Sinclair},\ and\ \citenamefont {Brener}}]{Campione15b}%
  \BibitemOpen
  \bibfield  {author} {\bibinfo {author} {\bibfnamefont {S.}~\bibnamefont
  {Campione}}, \bibinfo {author} {\bibfnamefont {S.}~\bibnamefont {Liu}},
  \bibinfo {author} {\bibfnamefont {A.}~\bibnamefont {Benz}}, \bibinfo {author}
  {\bibfnamefont {J.~F.}\ \bibnamefont {Klem}}, \bibinfo {author}
  {\bibfnamefont {M.~B.}\ \bibnamefont {Sinclair}},\ and\ \bibinfo {author}
  {\bibfnamefont {I.}~\bibnamefont {Brener}},\ }\bibfield  {title} {\bibinfo
  {title} {Epsilon-near-zero modes for tailored light-matter interaction},\
  }\href {https://doi.org/10.1103/PhysRevApplied.4.044011} {\bibfield
  {journal} {\bibinfo  {journal} {Phys. Rev. Appl.}\ }\textbf {\bibinfo
  {volume} {4}},\ \bibinfo {pages} {044011} (\bibinfo {year}
  {2015})}\BibitemShut {NoStop}%
\bibitem [{\citenamefont {Kinsey}\ \emph {et~al.}(2019)\citenamefont {Kinsey},
  \citenamefont {DeVault}, \citenamefont {Boltasseva},\ and\ \citenamefont
  {Shalaev}}]{kinsey19}%
  \BibitemOpen
  \bibfield  {author} {\bibinfo {author} {\bibfnamefont {N.}~\bibnamefont
  {Kinsey}}, \bibinfo {author} {\bibfnamefont {C.}~\bibnamefont {DeVault}},
  \bibinfo {author} {\bibfnamefont {A.}~\bibnamefont {Boltasseva}},\ and\
  \bibinfo {author} {\bibfnamefont {V.~M.}\ \bibnamefont {Shalaev}},\
  }\bibfield  {title} {\bibinfo {title} {Near-zero-index materials for
  photonics},\ }\href {https://doi.org/10.1038/s41578-019-0133-0} {\bibfield
  {journal} {\bibinfo  {journal} {Nature Reviews Materials}\ }\textbf {\bibinfo
  {volume} {4}},\ \bibinfo {pages} {742} (\bibinfo {year} {2019})}\BibitemShut
  {NoStop}%
\bibitem [{\citenamefont {Vesseur}\ \emph {et~al.}(2013)\citenamefont
  {Vesseur}, \citenamefont {Coenen}, \citenamefont {Caglayan}, \citenamefont
  {Engheta},\ and\ \citenamefont {Polman}}]{Vesseur13}%
  \BibitemOpen
  \bibfield  {author} {\bibinfo {author} {\bibfnamefont {E.~J.~R.}\
  \bibnamefont {Vesseur}}, \bibinfo {author} {\bibfnamefont {T.}~\bibnamefont
  {Coenen}}, \bibinfo {author} {\bibfnamefont {H.}~\bibnamefont {Caglayan}},
  \bibinfo {author} {\bibfnamefont {N.}~\bibnamefont {Engheta}},\ and\ \bibinfo
  {author} {\bibfnamefont {A.}~\bibnamefont {Polman}},\ }\bibfield  {title}
  {\bibinfo {title} {Experimental verification of $n=0$ structures for visible
  light},\ }\href {https://doi.org/10.1103/PhysRevLett.110.013902} {\bibfield
  {journal} {\bibinfo  {journal} {Phys. Rev. Lett.}\ }\textbf {\bibinfo
  {volume} {110}},\ \bibinfo {pages} {013902} (\bibinfo {year}
  {2013})}\BibitemShut {NoStop}%
\bibitem [{\citenamefont {Maas}\ \emph {et~al.}(2013)\citenamefont {Maas},
  \citenamefont {Parsons}, \citenamefont {Engheta},\ and\ \citenamefont
  {Polman}}]{Mass13}%
  \BibitemOpen
  \bibfield  {author} {\bibinfo {author} {\bibfnamefont {R.}~\bibnamefont
  {Maas}}, \bibinfo {author} {\bibfnamefont {J.}~\bibnamefont {Parsons}},
  \bibinfo {author} {\bibfnamefont {N.}~\bibnamefont {Engheta}},\ and\ \bibinfo
  {author} {\bibfnamefont {A.}~\bibnamefont {Polman}},\ }\bibfield  {title}
  {\bibinfo {title} {Experimental realization of an epsilon-near-zero
  metamaterial at visible wavelengths},\ }\href
  {https://doi.org/10.1038/nphoton.2013.256} {\bibfield  {journal} {\bibinfo
  {journal} {Nature Photonics}\ }\textbf {\bibinfo {volume} {7}},\ \bibinfo
  {pages} {907} (\bibinfo {year} {2013})}\BibitemShut {NoStop}%
\bibitem [{\citenamefont {Li}\ \emph {et~al.}(2015)\citenamefont {Li},
  \citenamefont {Kita}, \citenamefont {Mu{\~n}oz}, \citenamefont {Reshef},
  \citenamefont {Vulis}, \citenamefont {Yin}, \citenamefont {Lon{\v c}ar},\
  and\ \citenamefont {Mazur}}]{Li15}%
  \BibitemOpen
  \bibfield  {author} {\bibinfo {author} {\bibfnamefont {Y.}~\bibnamefont
  {Li}}, \bibinfo {author} {\bibfnamefont {S.}~\bibnamefont {Kita}}, \bibinfo
  {author} {\bibfnamefont {P.}~\bibnamefont {Mu{\~n}oz}}, \bibinfo {author}
  {\bibfnamefont {O.}~\bibnamefont {Reshef}}, \bibinfo {author} {\bibfnamefont
  {D.~I.}\ \bibnamefont {Vulis}}, \bibinfo {author} {\bibfnamefont
  {M.}~\bibnamefont {Yin}}, \bibinfo {author} {\bibfnamefont {M.}~\bibnamefont
  {Lon{\v c}ar}},\ and\ \bibinfo {author} {\bibfnamefont {E.}~\bibnamefont
  {Mazur}},\ }\bibfield  {title} {\bibinfo {title} {On-chip zero-index
  metamaterials},\ }\href {https://doi.org/10.1038/nphoton.2015.198} {\bibfield
   {journal} {\bibinfo  {journal} {Nature Photonics}\ }\textbf {\bibinfo
  {volume} {9}},\ \bibinfo {pages} {738} (\bibinfo {year} {2015})}\BibitemShut
  {NoStop}%
\bibitem [{\citenamefont {Huang}\ \emph {et~al.}(2011)\citenamefont {Huang},
  \citenamefont {Lai}, \citenamefont {Hang}, \citenamefont {Zheng},\ and\
  \citenamefont {Chan}}]{Huang11}%
  \BibitemOpen
  \bibfield  {author} {\bibinfo {author} {\bibfnamefont {X.}~\bibnamefont
  {Huang}}, \bibinfo {author} {\bibfnamefont {Y.}~\bibnamefont {Lai}}, \bibinfo
  {author} {\bibfnamefont {Z.~H.}\ \bibnamefont {Hang}}, \bibinfo {author}
  {\bibfnamefont {H.}~\bibnamefont {Zheng}},\ and\ \bibinfo {author}
  {\bibfnamefont {C.~T.}\ \bibnamefont {Chan}},\ }\bibfield  {title} {\bibinfo
  {title} {Dirac cones induced by accidental degeneracy in photonic crystals
  and zero-refractive-index materials},\ }\href
  {https://doi.org/10.1038/nmat3030} {\bibfield  {journal} {\bibinfo  {journal}
  {Nature Materials}\ }\textbf {\bibinfo {volume} {10}},\ \bibinfo {pages}
  {582} (\bibinfo {year} {2011})}\BibitemShut {NoStop}%
\bibitem [{\citenamefont {Chan}\ \emph {et~al.}(2012)\citenamefont {Chan},
  \citenamefont {Hang},\ and\ \citenamefont {Huang}}]{Chan12}%
  \BibitemOpen
  \bibfield  {author} {\bibinfo {author} {\bibfnamefont {C.~T.}\ \bibnamefont
  {Chan}}, \bibinfo {author} {\bibfnamefont {Z.~H.}\ \bibnamefont {Hang}},\
  and\ \bibinfo {author} {\bibfnamefont {X.}~\bibnamefont {Huang}},\ }\bibfield
   {title} {\bibinfo {title} {Dirac dispersion in two-dimensional photonic
  crystals},\ }\href {https://doi.org/https://doi.org/10.1155/2012/313984}
  {\bibfield  {journal} {\bibinfo  {journal} {Advances in OptoElectronics}\
  }\textbf {\bibinfo {volume} {2012}},\ \bibinfo {pages} {313984} (\bibinfo
  {year} {2012})}\BibitemShut {NoStop}%
\bibitem [{\citenamefont {Javanainen}\ \emph {et~al.}(1999)\citenamefont
  {Javanainen}, \citenamefont {Ruostekoski}, \citenamefont {Vestergaard},\ and\
  \citenamefont {Francis}}]{Javanainen1999a}%
  \BibitemOpen
  \bibfield  {author} {\bibinfo {author} {\bibfnamefont {J.}~\bibnamefont
  {Javanainen}}, \bibinfo {author} {\bibfnamefont {J.}~\bibnamefont
  {Ruostekoski}}, \bibinfo {author} {\bibfnamefont {B.}~\bibnamefont
  {Vestergaard}},\ and\ \bibinfo {author} {\bibfnamefont {M.~R.}\ \bibnamefont
  {Francis}},\ }\bibfield  {title} {\bibinfo {title} {One-dimensional modeling
  of light propagation in dense and degenerate samples},\ }\href
  {https://doi.org/10.1103/PhysRevA.59.649} {\bibfield  {journal} {\bibinfo
  {journal} {Phys. Rev. A}\ }\textbf {\bibinfo {volume} {59}},\ \bibinfo
  {pages} {649} (\bibinfo {year} {1999})}\BibitemShut {NoStop}%
\bibitem [{\citenamefont {Lee}\ \emph {et~al.}(2016)\citenamefont {Lee},
  \citenamefont {Jenkins},\ and\ \citenamefont {Ruostekoski}}]{Lee16}%
  \BibitemOpen
  \bibfield  {author} {\bibinfo {author} {\bibfnamefont {M.~D.}\ \bibnamefont
  {Lee}}, \bibinfo {author} {\bibfnamefont {S.~D.}\ \bibnamefont {Jenkins}},\
  and\ \bibinfo {author} {\bibfnamefont {J.}~\bibnamefont {Ruostekoski}},\
  }\bibfield  {title} {\bibinfo {title} {Stochastic methods for light
  propagation and recurrent scattering in saturated and nonsaturated atomic
  ensembles},\ }\href {https://doi.org/10.1103/PhysRevA.93.063803} {\bibfield
  {journal} {\bibinfo  {journal} {Phys. Rev. A}\ }\textbf {\bibinfo {volume}
  {93}},\ \bibinfo {pages} {063803} (\bibinfo {year} {2016})}\BibitemShut
  {NoStop}%
\bibitem [{\citenamefont {Rui}\ \emph {et~al.}(2020)\citenamefont {Rui},
  \citenamefont {Wei}, \citenamefont {Rubio-Abadal}, \citenamefont {Hollerith},
  \citenamefont {Zeiher}, \citenamefont {Stamper-Kurn}, \citenamefont {Gross},\
  and\ \citenamefont {Bloch}}]{Rui2020}%
  \BibitemOpen
  \bibfield  {author} {\bibinfo {author} {\bibfnamefont {J.}~\bibnamefont
  {Rui}}, \bibinfo {author} {\bibfnamefont {D.}~\bibnamefont {Wei}}, \bibinfo
  {author} {\bibfnamefont {A.}~\bibnamefont {Rubio-Abadal}}, \bibinfo {author}
  {\bibfnamefont {S.}~\bibnamefont {Hollerith}}, \bibinfo {author}
  {\bibfnamefont {J.}~\bibnamefont {Zeiher}}, \bibinfo {author} {\bibfnamefont
  {D.~M.}\ \bibnamefont {Stamper-Kurn}}, \bibinfo {author} {\bibfnamefont
  {C.}~\bibnamefont {Gross}},\ and\ \bibinfo {author} {\bibfnamefont
  {I.}~\bibnamefont {Bloch}},\ }\bibfield  {title} {\bibinfo {title} {{A
  subradiant optical mirror formed by a single structured atomic layer}},\
  }\href {https://doi.org/10.1038/s41586-020-2463-x} {\bibfield  {journal}
  {\bibinfo  {journal} {Nature}\ }\textbf {\bibinfo {volume} {583}},\ \bibinfo
  {pages} {369} (\bibinfo {year} {2020})}\BibitemShut {NoStop}%
\bibitem [{\citenamefont {Vatré}\ \emph {et~al.}(2024)\citenamefont {Vatré},
  \citenamefont {Lopes}, \citenamefont {Beugnon},\ and\ \citenamefont
  {Gerbier}}]{Vatre2024}%
  \BibitemOpen
  \bibfield  {author} {\bibinfo {author} {\bibfnamefont {R.}~\bibnamefont
  {Vatré}}, \bibinfo {author} {\bibfnamefont {R.}~\bibnamefont {Lopes}},
  \bibinfo {author} {\bibfnamefont {J.}~\bibnamefont {Beugnon}},\ and\ \bibinfo
  {author} {\bibfnamefont {F.}~\bibnamefont {Gerbier}},\ }\href
  {https://arxiv.org/abs/2409.04148} {\bibinfo {title} {Resonant light
  scattering by a slab of ultracold atoms}} (\bibinfo {year} {2024}),\ \Eprint
  {https://arxiv.org/abs/2409.04148} {arXiv:2409.04148 [cond-mat.quant-gas]}
  \BibitemShut {NoStop}%
\bibitem [{\citenamefont {Ruostekoski}(2023)}]{Ruostekoski2023}%
  \BibitemOpen
  \bibfield  {author} {\bibinfo {author} {\bibfnamefont {J.}~\bibnamefont
  {Ruostekoski}},\ }\bibfield  {title} {\bibinfo {title} {Cooperative
  quantum-optical planar arrays of atoms},\ }\href
  {https://doi.org/10.1103/PhysRevA.108.030101} {\bibfield  {journal} {\bibinfo
   {journal} {Phys. Rev. A}\ }\textbf {\bibinfo {volume} {108}},\ \bibinfo
  {pages} {030101} (\bibinfo {year} {2023})}\BibitemShut {NoStop}%
\bibitem [{\citenamefont {Reitz}\ \emph {et~al.}(2022)\citenamefont {Reitz},
  \citenamefont {Sommer},\ and\ \citenamefont {Genes}}]{Reitz22}%
  \BibitemOpen
  \bibfield  {author} {\bibinfo {author} {\bibfnamefont {M.}~\bibnamefont
  {Reitz}}, \bibinfo {author} {\bibfnamefont {C.}~\bibnamefont {Sommer}},\ and\
  \bibinfo {author} {\bibfnamefont {C.}~\bibnamefont {Genes}},\ }\bibfield
  {title} {\bibinfo {title} {Cooperative quantum phenomena in light-matter
  platforms},\ }\href {https://doi.org/10.1103/PRXQuantum.3.010201} {\bibfield
  {journal} {\bibinfo  {journal} {PRX Quantum}\ }\textbf {\bibinfo {volume}
  {3}},\ \bibinfo {pages} {010201} (\bibinfo {year} {2022})}\BibitemShut
  {NoStop}%
\bibitem [{\citenamefont {Shao}\ \emph {et~al.}(2024)\citenamefont {Shao},
  \citenamefont {Wang}, \citenamefont {Zhu}, \citenamefont {Zhu}, \citenamefont
  {Sun}, \citenamefont {Chen}, \citenamefont {Zhang}, \citenamefont {Fan},
  \citenamefont {Deng}, \citenamefont {Yao}, \citenamefont {Chen},\ and\
  \citenamefont {Pan}}]{Shao2024}%
  \BibitemOpen
  \bibfield  {author} {\bibinfo {author} {\bibfnamefont {H.-J.}\ \bibnamefont
  {Shao}}, \bibinfo {author} {\bibfnamefont {Y.-X.}\ \bibnamefont {Wang}},
  \bibinfo {author} {\bibfnamefont {D.-Z.}\ \bibnamefont {Zhu}}, \bibinfo
  {author} {\bibfnamefont {Y.-S.}\ \bibnamefont {Zhu}}, \bibinfo {author}
  {\bibfnamefont {H.-N.}\ \bibnamefont {Sun}}, \bibinfo {author} {\bibfnamefont
  {S.-Y.}\ \bibnamefont {Chen}}, \bibinfo {author} {\bibfnamefont
  {C.}~\bibnamefont {Zhang}}, \bibinfo {author} {\bibfnamefont {Z.-J.}\
  \bibnamefont {Fan}}, \bibinfo {author} {\bibfnamefont {Y.}~\bibnamefont
  {Deng}}, \bibinfo {author} {\bibfnamefont {X.-C.}\ \bibnamefont {Yao}},
  \bibinfo {author} {\bibfnamefont {Y.-A.}\ \bibnamefont {Chen}},\ and\
  \bibinfo {author} {\bibfnamefont {J.-W.}\ \bibnamefont {Pan}},\ }\bibfield
  {title} {\bibinfo {title} {Antiferromagnetic phase transition in a 3d
  fermionic hubbard model},\ }\href
  {https://doi.org/10.1038/s41586-024-07689-2} {\bibfield  {journal} {\bibinfo
  {journal} {Nature}\ }\textbf {\bibinfo {volume} {632}},\ \bibinfo {pages}
  {267} (\bibinfo {year} {2024})}\BibitemShut {NoStop}%
\bibitem [{\citenamefont {Srakaew}\ \emph {et~al.}(2023)\citenamefont
  {Srakaew}, \citenamefont {Weckesser}, \citenamefont {Hollerith},
  \citenamefont {Wei}, \citenamefont {Adler}, \citenamefont {Bloch},\ and\
  \citenamefont {Zeiher}}]{Srakaew22}%
  \BibitemOpen
  \bibfield  {author} {\bibinfo {author} {\bibfnamefont {K.}~\bibnamefont
  {Srakaew}}, \bibinfo {author} {\bibfnamefont {P.}~\bibnamefont {Weckesser}},
  \bibinfo {author} {\bibfnamefont {S.}~\bibnamefont {Hollerith}}, \bibinfo
  {author} {\bibfnamefont {D.}~\bibnamefont {Wei}}, \bibinfo {author}
  {\bibfnamefont {D.}~\bibnamefont {Adler}}, \bibinfo {author} {\bibfnamefont
  {I.}~\bibnamefont {Bloch}},\ and\ \bibinfo {author} {\bibfnamefont
  {J.}~\bibnamefont {Zeiher}},\ }\bibfield  {title} {\bibinfo {title} {A
  subwavelength atomic array switched by a single rydberg atom},\ }\href
  {https://doi.org/10.1038/s41567-023-01959-y} {\bibfield  {journal} {\bibinfo
  {journal} {Nature Physics}\ }\textbf {\bibinfo {volume} {19}},\ \bibinfo
  {pages} {714} (\bibinfo {year} {2023})}\BibitemShut {NoStop}%
\bibitem [{\citenamefont {Plankensteiner}\ \emph {et~al.}(2015)\citenamefont
  {Plankensteiner}, \citenamefont {Ostermann}, \citenamefont {Ritsch},\ and\
  \citenamefont {Genes}}]{Plankensteiner2015}%
  \BibitemOpen
  \bibfield  {author} {\bibinfo {author} {\bibfnamefont {D.}~\bibnamefont
  {Plankensteiner}}, \bibinfo {author} {\bibfnamefont {L.}~\bibnamefont
  {Ostermann}}, \bibinfo {author} {\bibfnamefont {H.}~\bibnamefont {Ritsch}},\
  and\ \bibinfo {author} {\bibfnamefont {C.}~\bibnamefont {Genes}},\ }\bibfield
   {title} {\bibinfo {title} {Selective protected state preparation of coupled
  dissipative quantum emitters},\ }\href {https://doi.org/10.1038/srep16231}
  {\bibfield  {journal} {\bibinfo  {journal} {Scientific Reports}\ }\textbf
  {\bibinfo {volume} {5}},\ \bibinfo {pages} {16231} (\bibinfo {year}
  {2015})}\BibitemShut {NoStop}%
\bibitem [{\citenamefont {Facchinetti}\ \emph {et~al.}(2016)\citenamefont
  {Facchinetti}, \citenamefont {Jenkins},\ and\ \citenamefont
  {Ruostekoski}}]{Facchinetti16}%
  \BibitemOpen
  \bibfield  {author} {\bibinfo {author} {\bibfnamefont {G.}~\bibnamefont
  {Facchinetti}}, \bibinfo {author} {\bibfnamefont {S.~D.}\ \bibnamefont
  {Jenkins}},\ and\ \bibinfo {author} {\bibfnamefont {J.}~\bibnamefont
  {Ruostekoski}},\ }\bibfield  {title} {\bibinfo {title} {Storing light with
  subradiant correlations in arrays of atoms},\ }\href
  {https://doi.org/10.1103/PhysRevLett.117.243601} {\bibfield  {journal}
  {\bibinfo  {journal} {Phys. Rev. Lett.}\ }\textbf {\bibinfo {volume} {117}},\
  \bibinfo {pages} {243601} (\bibinfo {year} {2016})}\BibitemShut {NoStop}%
\bibitem [{\citenamefont {Jen}\ \emph {et~al.}(2016)\citenamefont {Jen},
  \citenamefont {Chang},\ and\ \citenamefont {Chen}}]{Jen16}%
  \BibitemOpen
  \bibfield  {author} {\bibinfo {author} {\bibfnamefont {H.~H.}\ \bibnamefont
  {Jen}}, \bibinfo {author} {\bibfnamefont {M.-S.}\ \bibnamefont {Chang}},\
  and\ \bibinfo {author} {\bibfnamefont {Y.-C.}\ \bibnamefont {Chen}},\
  }\bibfield  {title} {\bibinfo {title} {Cooperative single-photon subradiant
  states},\ }\href {https://doi.org/10.1103/PhysRevA.94.013803} {\bibfield
  {journal} {\bibinfo  {journal} {Phys. Rev. A}\ }\textbf {\bibinfo {volume}
  {94}},\ \bibinfo {pages} {013803} (\bibinfo {year} {2016})}\BibitemShut
  {NoStop}%
\bibitem [{\citenamefont {Bettles}\ \emph {et~al.}(2015)\citenamefont
  {Bettles}, \citenamefont {Gardiner},\ and\ \citenamefont
  {Adams}}]{Bettles2015d}%
  \BibitemOpen
  \bibfield  {author} {\bibinfo {author} {\bibfnamefont {R.~J.}\ \bibnamefont
  {Bettles}}, \bibinfo {author} {\bibfnamefont {S.~A.}\ \bibnamefont
  {Gardiner}},\ and\ \bibinfo {author} {\bibfnamefont {C.~S.}\ \bibnamefont
  {Adams}},\ }\bibfield  {title} {\bibinfo {title} {{Cooperative ordering in
  lattices of interacting two-level dipoles}},\ }\href
  {https://doi.org/10.1103/PhysRevA.92.063822} {\bibfield  {journal} {\bibinfo
  {journal} {Phys. Rev. A}\ }\textbf {\bibinfo {volume} {92}},\ \bibinfo
  {pages} {063822} (\bibinfo {year} {2015})}\BibitemShut {NoStop}%
\bibitem [{\citenamefont {Sutherland}\ and\ \citenamefont
  {Robicheaux}(2016)}]{Sutherland1D}%
  \BibitemOpen
  \bibfield  {author} {\bibinfo {author} {\bibfnamefont {R.~T.}\ \bibnamefont
  {Sutherland}}\ and\ \bibinfo {author} {\bibfnamefont {F.}~\bibnamefont
  {Robicheaux}},\ }\bibfield  {title} {\bibinfo {title} {Collective
  dipole-dipole interactions in an atomic array},\ }\href
  {https://doi.org/10.1103/PhysRevA.94.013847} {\bibfield  {journal} {\bibinfo
  {journal} {Phys. Rev. A}\ }\textbf {\bibinfo {volume} {94}},\ \bibinfo
  {pages} {013847} (\bibinfo {year} {2016})}\BibitemShut {NoStop}%
\bibitem [{\citenamefont {Asenjo-Garcia}\ \emph {et~al.}(2017)\citenamefont
  {Asenjo-Garcia}, \citenamefont {Moreno-Cardoner}, \citenamefont {Albrecht},
  \citenamefont {Kimble},\ and\ \citenamefont {Chang}}]{Asenjo-Garcia2017a}%
  \BibitemOpen
  \bibfield  {author} {\bibinfo {author} {\bibfnamefont {A.}~\bibnamefont
  {Asenjo-Garcia}}, \bibinfo {author} {\bibfnamefont {M.}~\bibnamefont
  {Moreno-Cardoner}}, \bibinfo {author} {\bibfnamefont {A.}~\bibnamefont
  {Albrecht}}, \bibinfo {author} {\bibfnamefont {H.~J.}\ \bibnamefont
  {Kimble}},\ and\ \bibinfo {author} {\bibfnamefont {D.~E.}\ \bibnamefont
  {Chang}},\ }\bibfield  {title} {\bibinfo {title} {Exponential improvement in
  photon storage fidelities using subradiance and ``selective radiance'' in
  atomic arrays},\ }\href {https://doi.org/10.1103/PhysRevX.7.031024}
  {\bibfield  {journal} {\bibinfo  {journal} {Phys. Rev. X}\ }\textbf {\bibinfo
  {volume} {7}},\ \bibinfo {pages} {031024} (\bibinfo {year}
  {2017})}\BibitemShut {NoStop}%
\bibitem [{\citenamefont {Guimond}\ \emph {et~al.}(2019)\citenamefont
  {Guimond}, \citenamefont {Grankin}, \citenamefont {Vasilyev}, \citenamefont
  {Vermersch},\ and\ \citenamefont {Zoller}}]{Guimond2019}%
  \BibitemOpen
  \bibfield  {author} {\bibinfo {author} {\bibfnamefont {P.-O.}\ \bibnamefont
  {Guimond}}, \bibinfo {author} {\bibfnamefont {A.}~\bibnamefont {Grankin}},
  \bibinfo {author} {\bibfnamefont {D.~V.}\ \bibnamefont {Vasilyev}}, \bibinfo
  {author} {\bibfnamefont {B.}~\bibnamefont {Vermersch}},\ and\ \bibinfo
  {author} {\bibfnamefont {P.}~\bibnamefont {Zoller}},\ }\bibfield  {title}
  {\bibinfo {title} {Subradiant {B}ell states in distant atomic arrays},\
  }\href {https://doi.org/10.1103/PhysRevLett.122.093601} {\bibfield  {journal}
  {\bibinfo  {journal} {Phys. Rev. Lett.}\ }\textbf {\bibinfo {volume} {122}},\
  \bibinfo {pages} {093601} (\bibinfo {year} {2019})}\BibitemShut {NoStop}%
\bibitem [{\citenamefont {Ballantine}\ and\ \citenamefont
  {Ruostekoski}(2020)}]{Ballantine20ant}%
  \BibitemOpen
  \bibfield  {author} {\bibinfo {author} {\bibfnamefont {K.~E.}\ \bibnamefont
  {Ballantine}}\ and\ \bibinfo {author} {\bibfnamefont {J.}~\bibnamefont
  {Ruostekoski}},\ }\bibfield  {title} {\bibinfo {title} {Subradiance-protected
  excitation spreading in the generation of collimated photon emission from an
  atomic array},\ }\href {https://doi.org/10.1103/PhysRevResearch.2.023086}
  {\bibfield  {journal} {\bibinfo  {journal} {Phys. Rev. Research}\ }\textbf
  {\bibinfo {volume} {2}},\ \bibinfo {pages} {023086} (\bibinfo {year}
  {2020})}\BibitemShut {NoStop}%
\bibitem [{\citenamefont {Shah}\ \emph {et~al.}(2024)\citenamefont {Shah},
  \citenamefont {Patti}, \citenamefont {Rubies-Bigorda},\ and\ \citenamefont
  {Yelin}}]{Shah24}%
  \BibitemOpen
  \bibfield  {author} {\bibinfo {author} {\bibfnamefont {F.}~\bibnamefont
  {Shah}}, \bibinfo {author} {\bibfnamefont {T.~L.}\ \bibnamefont {Patti}},
  \bibinfo {author} {\bibfnamefont {O.}~\bibnamefont {Rubies-Bigorda}},\ and\
  \bibinfo {author} {\bibfnamefont {S.~F.}\ \bibnamefont {Yelin}},\ }\bibfield
  {title} {\bibinfo {title} {Quantum computing with subwavelength atomic
  arrays},\ }\href {https://doi.org/10.1103/PhysRevA.109.012613} {\bibfield
  {journal} {\bibinfo  {journal} {Phys. Rev. A}\ }\textbf {\bibinfo {volume}
  {109}},\ \bibinfo {pages} {012613} (\bibinfo {year} {2024})}\BibitemShut
  {NoStop}%
\bibitem [{\citenamefont {Plankensteiner}\ \emph {et~al.}(2017)\citenamefont
  {Plankensteiner}, \citenamefont {Sommer}, \citenamefont {Ritsch},\ and\
  \citenamefont {Genes}}]{Plankensteiner2017}%
  \BibitemOpen
  \bibfield  {author} {\bibinfo {author} {\bibfnamefont {D.}~\bibnamefont
  {Plankensteiner}}, \bibinfo {author} {\bibfnamefont {C.}~\bibnamefont
  {Sommer}}, \bibinfo {author} {\bibfnamefont {H.}~\bibnamefont {Ritsch}},\
  and\ \bibinfo {author} {\bibfnamefont {C.}~\bibnamefont {Genes}},\ }\bibfield
   {title} {\bibinfo {title} {Cavity antiresonance spectroscopy of dipole
  coupled subradiant arrays},\ }\href
  {https://doi.org/10.1103/PhysRevLett.119.093601} {\bibfield  {journal}
  {\bibinfo  {journal} {Phys. Rev. Lett.}\ }\textbf {\bibinfo {volume} {119}},\
  \bibinfo {pages} {093601} (\bibinfo {year} {2017})}\BibitemShut {NoStop}%
\bibitem [{\citenamefont {Parmee}\ and\ \citenamefont
  {Ruostekoski}(2021)}]{Parmee20bistable}%
  \BibitemOpen
  \bibfield  {author} {\bibinfo {author} {\bibfnamefont {C.~D.}\ \bibnamefont
  {Parmee}}\ and\ \bibinfo {author} {\bibfnamefont {J.}~\bibnamefont
  {Ruostekoski}},\ }\bibfield  {title} {\bibinfo {title} {Bistable optical
  transmission through arrays of atoms in free space},\ }\href
  {https://doi.org/10.1103/PhysRevA.103.033706} {\bibfield  {journal} {\bibinfo
   {journal} {Phys. Rev. A}\ }\textbf {\bibinfo {volume} {103}},\ \bibinfo
  {pages} {033706} (\bibinfo {year} {2021})}\BibitemShut {NoStop}%
\bibitem [{\citenamefont {Pedersen}\ \emph {et~al.}(2023)\citenamefont
  {Pedersen}, \citenamefont {Zhang},\ and\ \citenamefont {Pohl}}]{Pedersen23}%
  \BibitemOpen
  \bibfield  {author} {\bibinfo {author} {\bibfnamefont {S.~P.}\ \bibnamefont
  {Pedersen}}, \bibinfo {author} {\bibfnamefont {L.}~\bibnamefont {Zhang}},\
  and\ \bibinfo {author} {\bibfnamefont {T.}~\bibnamefont {Pohl}},\ }\bibfield
  {title} {\bibinfo {title} {Quantum nonlinear metasurfaces from dual arrays of
  ultracold atoms},\ }\href {https://doi.org/10.1103/PhysRevResearch.5.L012047}
  {\bibfield  {journal} {\bibinfo  {journal} {Phys. Rev. Res.}\ }\textbf
  {\bibinfo {volume} {5}},\ \bibinfo {pages} {L012047} (\bibinfo {year}
  {2023})}\BibitemShut {NoStop}%
\bibitem [{\citenamefont {Castells-Graells}\ \emph {et~al.}(2025)\citenamefont
  {Castells-Graells}, \citenamefont {Cirac},\ and\ \citenamefont
  {Wild}}]{Castells25}%
  \BibitemOpen
  \bibfield  {author} {\bibinfo {author} {\bibfnamefont {D.}~\bibnamefont
  {Castells-Graells}}, \bibinfo {author} {\bibfnamefont {J.~I.}\ \bibnamefont
  {Cirac}},\ and\ \bibinfo {author} {\bibfnamefont {D.~S.}\ \bibnamefont
  {Wild}},\ }\bibfield  {title} {\bibinfo {title} {Cavity quantum
  electrodynamics with atom arrays in free space},\ }\href
  {https://doi.org/10.1103/PhysRevA.111.053712} {\bibfield  {journal} {\bibinfo
   {journal} {Phys. Rev. A}\ }\textbf {\bibinfo {volume} {111}},\ \bibinfo
  {pages} {053712} (\bibinfo {year} {2025})}\BibitemShut {NoStop}%
\bibitem [{\citenamefont {Sinha}\ \emph {et~al.}(2025)\citenamefont {Sinha},
  \citenamefont {Parra-Contreras}, \citenamefont {Das},\ and\ \citenamefont
  {Solano}}]{Sinha2025}%
  \BibitemOpen
  \bibfield  {author} {\bibinfo {author} {\bibfnamefont {K.}~\bibnamefont
  {Sinha}}, \bibinfo {author} {\bibfnamefont {J.}~\bibnamefont
  {Parra-Contreras}}, \bibinfo {author} {\bibfnamefont {A.}~\bibnamefont
  {Das}},\ and\ \bibinfo {author} {\bibfnamefont {P.}~\bibnamefont {Solano}},\
  }\bibfield  {title} {\bibinfo {title} {Spontaneous emission in the presence
  of quantum mirrors},\ }\href {https://doi.org/10.1088/1367-2630/add495}
  {\bibfield  {journal} {\bibinfo  {journal} {New Journal of Physics}\ }\textbf
  {\bibinfo {volume} {27}},\ \bibinfo {pages} {054101} (\bibinfo {year}
  {2025})}\BibitemShut {NoStop}%
\bibitem [{\citenamefont {Bekenstein}\ \emph {et~al.}(2020)\citenamefont
  {Bekenstein}, \citenamefont {Pikovski}, \citenamefont {Pichler},
  \citenamefont {Shahmoon}, \citenamefont {Yelin},\ and\ \citenamefont
  {Lukin}}]{Bekenstein2020}%
  \BibitemOpen
  \bibfield  {author} {\bibinfo {author} {\bibfnamefont {R.}~\bibnamefont
  {Bekenstein}}, \bibinfo {author} {\bibfnamefont {I.}~\bibnamefont
  {Pikovski}}, \bibinfo {author} {\bibfnamefont {H.}~\bibnamefont {Pichler}},
  \bibinfo {author} {\bibfnamefont {E.}~\bibnamefont {Shahmoon}}, \bibinfo
  {author} {\bibfnamefont {S.~F.}\ \bibnamefont {Yelin}},\ and\ \bibinfo
  {author} {\bibfnamefont {M.~D.}\ \bibnamefont {Lukin}},\ }\bibfield  {title}
  {\bibinfo {title} {Quantum metasurfaces with atom arrays},\ }\href
  {https://doi.org/10.1038/s41567-020-0845-5} {\bibfield  {journal} {\bibinfo
  {journal} {Nature Physics}\ }\textbf {\bibinfo {volume} {16}},\ \bibinfo
  {pages} {676} (\bibinfo {year} {2020})}\BibitemShut {NoStop}%
\bibitem [{\citenamefont {Moreno-Cardoner}\ \emph {et~al.}(2021)\citenamefont
  {Moreno-Cardoner}, \citenamefont {Goncalves},\ and\ \citenamefont
  {Chang}}]{Cardoner21}%
  \BibitemOpen
  \bibfield  {author} {\bibinfo {author} {\bibfnamefont {M.}~\bibnamefont
  {Moreno-Cardoner}}, \bibinfo {author} {\bibfnamefont {D.}~\bibnamefont
  {Goncalves}},\ and\ \bibinfo {author} {\bibfnamefont {D.~E.}\ \bibnamefont
  {Chang}},\ }\bibfield  {title} {\bibinfo {title} {Quantum nonlinear optics
  based on two-dimensional rydberg atom arrays},\ }\href
  {https://doi.org/10.1103/PhysRevLett.127.263602} {\bibfield  {journal}
  {\bibinfo  {journal} {Phys. Rev. Lett.}\ }\textbf {\bibinfo {volume} {127}},\
  \bibinfo {pages} {263602} (\bibinfo {year} {2021})}\BibitemShut {NoStop}%
\bibitem [{\citenamefont {Zhang}\ \emph {et~al.}(2022)\citenamefont {Zhang},
  \citenamefont {Walther}, \citenamefont {M{\o{}}lmer},\ and\ \citenamefont
  {Pohl}}]{Zhang2022}%
  \BibitemOpen
  \bibfield  {author} {\bibinfo {author} {\bibfnamefont {L.}~\bibnamefont
  {Zhang}}, \bibinfo {author} {\bibfnamefont {V.}~\bibnamefont {Walther}},
  \bibinfo {author} {\bibfnamefont {K.}~\bibnamefont {M{\o{}}lmer}},\ and\
  \bibinfo {author} {\bibfnamefont {T.}~\bibnamefont {Pohl}},\ }\bibfield
  {title} {\bibinfo {title} {Photon-photon interactions in {R}ydberg-atom
  arrays},\ }\href {https://doi.org/10.22331/q-2022-03-30-674} {\bibfield
  {journal} {\bibinfo  {journal} {{Quantum}}\ }\textbf {\bibinfo {volume}
  {6}},\ \bibinfo {pages} {674} (\bibinfo {year} {2022})}\BibitemShut {NoStop}%
\bibitem [{\citenamefont {Ruostekoski}\ and\ \citenamefont
  {Javanainen}(1997)}]{Ruostekoski1997a}%
  \BibitemOpen
  \bibfield  {author} {\bibinfo {author} {\bibfnamefont {J.}~\bibnamefont
  {Ruostekoski}}\ and\ \bibinfo {author} {\bibfnamefont {J.}~\bibnamefont
  {Javanainen}},\ }\bibfield  {title} {\bibinfo {title} {Quantum field theory
  of cooperative atom response: Low light intensity},\ }\href
  {https://doi.org/10.1103/PhysRevA.55.513} {\bibfield  {journal} {\bibinfo
  {journal} {Phys. Rev. A}\ }\textbf {\bibinfo {volume} {55}},\ \bibinfo
  {pages} {513} (\bibinfo {year} {1997})}\BibitemShut {NoStop}%
\bibitem [{\citenamefont {Lax}(1951)}]{Lax51}%
  \BibitemOpen
  \bibfield  {author} {\bibinfo {author} {\bibfnamefont {M.}~\bibnamefont
  {Lax}},\ }\bibfield  {title} {\bibinfo {title} {Multiple scattering of
  waves},\ }\href {https://doi.org/10.1103/RevModPhys.23.287} {\bibfield
  {journal} {\bibinfo  {journal} {Rev. Mod. Phys.}\ }\textbf {\bibinfo {volume}
  {23}},\ \bibinfo {pages} {287} (\bibinfo {year} {1951})}\BibitemShut
  {NoStop}%
\bibitem [{\citenamefont {Morice}\ \emph {et~al.}(1995)\citenamefont {Morice},
  \citenamefont {Castin},\ and\ \citenamefont {Dalibard}}]{Morice1995a}%
  \BibitemOpen
  \bibfield  {author} {\bibinfo {author} {\bibfnamefont {O.}~\bibnamefont
  {Morice}}, \bibinfo {author} {\bibfnamefont {Y.}~\bibnamefont {Castin}},\
  and\ \bibinfo {author} {\bibfnamefont {J.}~\bibnamefont {Dalibard}},\
  }\bibfield  {title} {\bibinfo {title} {Refractive index of a dilute {B}ose
  gas},\ }\href {https://doi.org/10.1103/PhysRevA.51.3896} {\bibfield
  {journal} {\bibinfo  {journal} {Phys. Rev. A}\ }\textbf {\bibinfo {volume}
  {51}},\ \bibinfo {pages} {3896} (\bibinfo {year} {1995})}\BibitemShut
  {NoStop}%
\bibitem [{\citenamefont {Sokolov}\ \emph {et~al.}(2011)\citenamefont
  {Sokolov}, \citenamefont {Kupriyanov},\ and\ \citenamefont
  {Havey}}]{Sokolov2011}%
  \BibitemOpen
  \bibfield  {author} {\bibinfo {author} {\bibfnamefont {I.~M.}\ \bibnamefont
  {Sokolov}}, \bibinfo {author} {\bibfnamefont {D.~V.}\ \bibnamefont
  {Kupriyanov}},\ and\ \bibinfo {author} {\bibfnamefont {M.~D.}\ \bibnamefont
  {Havey}},\ }\bibfield  {title} {\bibinfo {title} {Microscopic theory of
  scattering of weak electromagnetic radiation by a dense ensemble of ultracold
  atoms},\ }\href {https://doi.org/10.1134/S106377611101016X} {\bibfield
  {journal} {\bibinfo  {journal} {Journal of Experimental and Theoretical
  Physics}\ }\textbf {\bibinfo {volume} {112}},\ \bibinfo {pages} {246}
  (\bibinfo {year} {2011})}\BibitemShut {NoStop}%
\bibitem [{\citenamefont {Ruks}\ \emph {et~al.}(2025)\citenamefont {Ruks},
  \citenamefont {Ballantine},\ and\ \citenamefont {Ruostekoski}}]{ruks25}%
  \BibitemOpen
  \bibfield  {author} {\bibinfo {author} {\bibfnamefont {L.}~\bibnamefont
  {Ruks}}, \bibinfo {author} {\bibfnamefont {K.~E.}\ \bibnamefont
  {Ballantine}},\ and\ \bibinfo {author} {\bibfnamefont {J.}~\bibnamefont
  {Ruostekoski}},\ }\bibfield  {title} {\bibinfo {title} {Negative refraction
  of light in an atomic medium},\ }\href
  {https://doi.org/10.1038/s41467-025-56250-w} {\bibfield  {journal} {\bibinfo
  {journal} {Nature Communications}\ }\textbf {\bibinfo {volume} {16}},\
  \bibinfo {pages} {1433} (\bibinfo {year} {2025})}\BibitemShut {NoStop}%
\bibitem [{\citenamefont {Antezza}\ and\ \citenamefont
  {Castin}(2009{\natexlab{a}})}]{castin09}%
  \BibitemOpen
  \bibfield  {author} {\bibinfo {author} {\bibfnamefont {M.}~\bibnamefont
  {Antezza}}\ and\ \bibinfo {author} {\bibfnamefont {Y.}~\bibnamefont
  {Castin}},\ }\bibfield  {title} {\bibinfo {title} {Spectrum of light in a
  quantum fluctuating periodic structure},\ }\href
  {https://doi.org/10.1103/PhysRevLett.103.123903} {\bibfield  {journal}
  {\bibinfo  {journal} {Phys. Rev. Lett.}\ }\textbf {\bibinfo {volume} {103}},\
  \bibinfo {pages} {123903} (\bibinfo {year} {2009}{\natexlab{a}})}\BibitemShut
  {NoStop}%
\bibitem [{\citenamefont {Antezza}\ and\ \citenamefont
  {Castin}(2009{\natexlab{b}})}]{Antezza09b}%
  \BibitemOpen
  \bibfield  {author} {\bibinfo {author} {\bibfnamefont {M.}~\bibnamefont
  {Antezza}}\ and\ \bibinfo {author} {\bibfnamefont {Y.}~\bibnamefont
  {Castin}},\ }\bibfield  {title} {\bibinfo {title} {Fano-hopfield model and
  photonic band gaps for an arbitrary atomic lattice},\ }\href
  {https://doi.org/10.1103/PhysRevA.80.013816} {\bibfield  {journal} {\bibinfo
  {journal} {Phys. Rev. A}\ }\textbf {\bibinfo {volume} {80}},\ \bibinfo
  {pages} {013816} (\bibinfo {year} {2009}{\natexlab{b}})}\BibitemShut
  {NoStop}%
\bibitem [{\citenamefont {Popov}\ \emph {et~al.}(2019)\citenamefont {Popov},
  \citenamefont {Tretyakov},\ and\ \citenamefont {Novitsky}}]{Popov2019}%
  \BibitemOpen
  \bibfield  {author} {\bibinfo {author} {\bibfnamefont {V.}~\bibnamefont
  {Popov}}, \bibinfo {author} {\bibfnamefont {S.}~\bibnamefont {Tretyakov}},\
  and\ \bibinfo {author} {\bibfnamefont {A.}~\bibnamefont {Novitsky}},\
  }\bibfield  {title} {\bibinfo {title} {Brewster effect when approaching
  exceptional points of degeneracy: Epsilon-near-zero behavior},\ }\href
  {https://doi.org/10.1103/PhysRevB.99.045146} {\bibfield  {journal} {\bibinfo
  {journal} {Phys. Rev. B}\ }\textbf {\bibinfo {volume} {99}},\ \bibinfo
  {pages} {045146} (\bibinfo {year} {2019})}\BibitemShut {NoStop}%
\bibitem [{\citenamefont {Javanainen}\ \emph {et~al.}(2017)\citenamefont
  {Javanainen}, \citenamefont {Ruostekoski}, \citenamefont {Li},\ and\
  \citenamefont {Yoo}}]{Javanainen17}%
  \BibitemOpen
  \bibfield  {author} {\bibinfo {author} {\bibfnamefont {J.}~\bibnamefont
  {Javanainen}}, \bibinfo {author} {\bibfnamefont {J.}~\bibnamefont
  {Ruostekoski}}, \bibinfo {author} {\bibfnamefont {Y.}~\bibnamefont {Li}},\
  and\ \bibinfo {author} {\bibfnamefont {S.-M.}\ \bibnamefont {Yoo}},\
  }\bibfield  {title} {\bibinfo {title} {Exact electrodynamics versus standard
  optics for a slab of cold dense gas},\ }\href
  {https://doi.org/10.1103/PhysRevA.96.033835} {\bibfield  {journal} {\bibinfo
  {journal} {Phys. Rev. A}\ }\textbf {\bibinfo {volume} {96}},\ \bibinfo
  {pages} {033835} (\bibinfo {year} {2017})}\BibitemShut {NoStop}%
\bibitem [{\citenamefont {McCutcheon}\ \emph {et~al.}(2024)\citenamefont
  {McCutcheon}, \citenamefont {Ostermann},\ and\ \citenamefont
  {Yelin}}]{McCutcheon24}%
  \BibitemOpen
  \bibfield  {author} {\bibinfo {author} {\bibfnamefont {R.~A.}\ \bibnamefont
  {McCutcheon}}, \bibinfo {author} {\bibfnamefont {S.}~\bibnamefont
  {Ostermann}},\ and\ \bibinfo {author} {\bibfnamefont {S.~F.}\ \bibnamefont
  {Yelin}},\ }\bibfield  {title} {\bibinfo {title} {Effect of photon
  propagation on a near-zero-refractive-index medium},\ }\href
  {https://doi.org/10.1103/PhysRevA.110.033705} {\bibfield  {journal} {\bibinfo
   {journal} {Phys. Rev. A}\ }\textbf {\bibinfo {volume} {110}},\ \bibinfo
  {pages} {033705} (\bibinfo {year} {2024})}\BibitemShut {NoStop}%
\bibitem [{\citenamefont {Smith}\ \emph {et~al.}(2002)\citenamefont {Smith},
  \citenamefont {Schultz}, \citenamefont {Marko\ifmmode~\check{s}\else
  \v{s}\fi{}},\ and\ \citenamefont {Soukoulis}}]{SmithEtAlPRB2002}%
  \BibitemOpen
  \bibfield  {author} {\bibinfo {author} {\bibfnamefont {D.~R.}\ \bibnamefont
  {Smith}}, \bibinfo {author} {\bibfnamefont {S.}~\bibnamefont {Schultz}},
  \bibinfo {author} {\bibfnamefont {P.}~\bibnamefont
  {Marko\ifmmode~\check{s}\else \v{s}\fi{}}},\ and\ \bibinfo {author}
  {\bibfnamefont {C.~M.}\ \bibnamefont {Soukoulis}},\ }\bibfield  {title}
  {\bibinfo {title} {Determination of effective permittivity and permeability
  of metamaterials from reflection and transmission coefficients},\ }\href
  {https://doi.org/10.1103/PhysRevB.65.195104} {\bibfield  {journal} {\bibinfo
  {journal} {Phys. Rev. B}\ }\textbf {\bibinfo {volume} {65}},\ \bibinfo
  {pages} {195104} (\bibinfo {year} {2002})}\BibitemShut {NoStop}%
\bibitem [{\citenamefont {Al\`u}(2011)}]{Alu2011}%
  \BibitemOpen
  \bibfield  {author} {\bibinfo {author} {\bibfnamefont {A.}~\bibnamefont
  {Al\`u}},\ }\bibfield  {title} {\bibinfo {title} {First-principles
  homogenization theory for periodic metamaterials},\ }\href
  {https://doi.org/10.1103/PhysRevB.84.075153} {\bibfield  {journal} {\bibinfo
  {journal} {Phys. Rev. B}\ }\textbf {\bibinfo {volume} {84}},\ \bibinfo
  {pages} {075153} (\bibinfo {year} {2011})}\BibitemShut {NoStop}%
\bibitem [{\citenamefont {Jenkins}\ and\ \citenamefont
  {Ruostekoski}(2012)}]{Jenkins2012a}%
  \BibitemOpen
  \bibfield  {author} {\bibinfo {author} {\bibfnamefont {S.~D.}\ \bibnamefont
  {Jenkins}}\ and\ \bibinfo {author} {\bibfnamefont {J.}~\bibnamefont
  {Ruostekoski}},\ }\bibfield  {title} {\bibinfo {title} {Controlled
  manipulation of light by cooperative response of atoms in an optical
  lattice},\ }\href {https://doi.org/10.1103/PhysRevA.86.031602} {\bibfield
  {journal} {\bibinfo  {journal} {Phys. Rev. A}\ }\textbf {\bibinfo {volume}
  {86}},\ \bibinfo {pages} {031602} (\bibinfo {year} {2012})}\BibitemShut
  {NoStop}%
\bibitem [{\citenamefont {Dicke}(1954)}]{Dicke54}%
  \BibitemOpen
  \bibfield  {author} {\bibinfo {author} {\bibfnamefont {R.~H.}\ \bibnamefont
  {Dicke}},\ }\bibfield  {title} {\bibinfo {title} {Coherence in spontaneous
  radiation processes},\ }\href {https://doi.org/10.1103/PhysRev.93.99}
  {\bibfield  {journal} {\bibinfo  {journal} {Phys. Rev.}\ }\textbf {\bibinfo
  {volume} {93}},\ \bibinfo {pages} {99} (\bibinfo {year} {1954})}\BibitemShut
  {NoStop}%
\bibitem [{\citenamefont {Walls}\ \emph {et~al.}(1978)\citenamefont {Walls},
  \citenamefont {Drummond}, \citenamefont {Hassan},\ and\ \citenamefont
  {Carmichael}}]{Walls1978}%
  \BibitemOpen
  \bibfield  {author} {\bibinfo {author} {\bibfnamefont {D.~F.}\ \bibnamefont
  {Walls}}, \bibinfo {author} {\bibfnamefont {P.~D.}\ \bibnamefont {Drummond}},
  \bibinfo {author} {\bibfnamefont {S.~S.}\ \bibnamefont {Hassan}},\ and\
  \bibinfo {author} {\bibfnamefont {H.~J.}\ \bibnamefont {Carmichael}},\
  }\bibfield  {title} {\bibinfo {title} {{Non-Equilibrium Phase Transitions in
  Cooperative Atomic Systems}},\ }\href {https://doi.org/10.1143/PTPS.64.307}
  {\bibfield  {journal} {\bibinfo  {journal} {Progress of Theoretical Physics
  Supplement}\ }\textbf {\bibinfo {volume} {64}},\ \bibinfo {pages} {307}
  (\bibinfo {year} {1978})}\BibitemShut {NoStop}%
\bibitem [{\citenamefont {Carmichael}(1980)}]{Carmichael1980}%
  \BibitemOpen
  \bibfield  {author} {\bibinfo {author} {\bibfnamefont {H.~J.}\ \bibnamefont
  {Carmichael}},\ }\bibfield  {title} {\bibinfo {title} {Analytical and
  numerical results for the steady state in cooperative resonance
  fluorescence},\ }\href {https://doi.org/10.1088/0022-3700/13/18/009}
  {\bibfield  {journal} {\bibinfo  {journal} {Journal of Physics B: Atomic and
  Molecular Physics}\ }\textbf {\bibinfo {volume} {13}},\ \bibinfo {pages}
  {3551} (\bibinfo {year} {1980})}\BibitemShut {NoStop}%
\bibitem [{\citenamefont {Ferioli}\ \emph {et~al.}(2023)\citenamefont
  {Ferioli}, \citenamefont {Glicenstein}, \citenamefont {Ferrier-Barbut},\ and\
  \citenamefont {Browaeys}}]{Ferioli2023}%
  \BibitemOpen
  \bibfield  {author} {\bibinfo {author} {\bibfnamefont {G.}~\bibnamefont
  {Ferioli}}, \bibinfo {author} {\bibfnamefont {A.}~\bibnamefont
  {Glicenstein}}, \bibinfo {author} {\bibfnamefont {I.}~\bibnamefont
  {Ferrier-Barbut}},\ and\ \bibinfo {author} {\bibfnamefont {A.}~\bibnamefont
  {Browaeys}},\ }\bibfield  {title} {\bibinfo {title} {A non-equilibrium
  superradiant phase transition in free space},\ }\href
  {https://doi.org/10.1038/s41567-023-02064-w} {\bibfield  {journal} {\bibinfo
  {journal} {Nature Physics}\ }\textbf {\bibinfo {volume} {19}},\ \bibinfo
  {pages} {1345} (\bibinfo {year} {2023})}\BibitemShut {NoStop}%
\bibitem [{\citenamefont {Agarwal}\ \emph {et~al.}(2024)\citenamefont
  {Agarwal}, \citenamefont {Chaparro}, \citenamefont {Barberena}, \citenamefont
  {Orioli}, \citenamefont {Ferioli}, \citenamefont {Pancaldi}, \citenamefont
  {Ferrier-Barbut}, \citenamefont {Browaeys},\ and\ \citenamefont
  {Rey}}]{Agarwal2024}%
  \BibitemOpen
  \bibfield  {author} {\bibinfo {author} {\bibfnamefont {S.}~\bibnamefont
  {Agarwal}}, \bibinfo {author} {\bibfnamefont {E.}~\bibnamefont {Chaparro}},
  \bibinfo {author} {\bibfnamefont {D.}~\bibnamefont {Barberena}}, \bibinfo
  {author} {\bibfnamefont {A.~P.~n.}\ \bibnamefont {Orioli}}, \bibinfo {author}
  {\bibfnamefont {G.}~\bibnamefont {Ferioli}}, \bibinfo {author} {\bibfnamefont
  {S.}~\bibnamefont {Pancaldi}}, \bibinfo {author} {\bibfnamefont
  {I.}~\bibnamefont {Ferrier-Barbut}}, \bibinfo {author} {\bibfnamefont
  {A.}~\bibnamefont {Browaeys}},\ and\ \bibinfo {author} {\bibfnamefont
  {A.}~\bibnamefont {Rey}},\ }\bibfield  {title} {\bibinfo {title} {Directional
  superradiance in a driven ultracold atomic gas in free space},\ }\href
  {https://doi.org/10.1103/PRXQuantum.5.040335} {\bibfield  {journal} {\bibinfo
   {journal} {PRX Quantum}\ }\textbf {\bibinfo {volume} {5}},\ \bibinfo {pages}
  {040335} (\bibinfo {year} {2024})}\BibitemShut {NoStop}%
\bibitem [{\citenamefont {Ruostekoski}(2025)}]{Ruostekoski25}%
  \BibitemOpen
  \bibfield  {author} {\bibinfo {author} {\bibfnamefont {J.}~\bibnamefont
  {Ruostekoski}},\ }\bibfield  {title} {\bibinfo {title} {Superradiant phase
  transition in a large interacting driven atomic ensemble in free space},\
  }\href {https://doi.org/10.1364/OPTICAQ.537927} {\bibfield  {journal}
  {\bibinfo  {journal} {Optica Quantum}\ }\textbf {\bibinfo {volume} {3}},\
  \bibinfo {pages} {15} (\bibinfo {year} {2025})}\BibitemShut {NoStop}%
\bibitem [{\citenamefont {Goncalves}\ \emph {et~al.}(2025)\citenamefont
  {Goncalves}, \citenamefont {Bombieri}, \citenamefont {Ferioli}, \citenamefont
  {Pancaldi}, \citenamefont {Ferrier-Barbut}, \citenamefont {Browaeys},
  \citenamefont {Shahmoon},\ and\ \citenamefont {Chang}}]{Goncalves25}%
  \BibitemOpen
  \bibfield  {author} {\bibinfo {author} {\bibfnamefont {D.}~\bibnamefont
  {Goncalves}}, \bibinfo {author} {\bibfnamefont {L.}~\bibnamefont {Bombieri}},
  \bibinfo {author} {\bibfnamefont {G.}~\bibnamefont {Ferioli}}, \bibinfo
  {author} {\bibfnamefont {S.}~\bibnamefont {Pancaldi}}, \bibinfo {author}
  {\bibfnamefont {I.}~\bibnamefont {Ferrier-Barbut}}, \bibinfo {author}
  {\bibfnamefont {A.}~\bibnamefont {Browaeys}}, \bibinfo {author}
  {\bibfnamefont {E.}~\bibnamefont {Shahmoon}},\ and\ \bibinfo {author}
  {\bibfnamefont {D.}~\bibnamefont {Chang}},\ }\bibfield  {title} {\bibinfo
  {title} {Driven-dissipative phase separation in free-space atomic
  ensembles},\ }\href {https://doi.org/10.1103/PRXQuantum.6.020303} {\bibfield
  {journal} {\bibinfo  {journal} {PRX Quantum}\ }\textbf {\bibinfo {volume}
  {6}},\ \bibinfo {pages} {020303} (\bibinfo {year} {2025})}\BibitemShut
  {NoStop}%
\bibitem [{\citenamefont {Olmos}\ \emph {et~al.}(2013)\citenamefont {Olmos},
  \citenamefont {Yu}, \citenamefont {Singh}, \citenamefont {Schreck},
  \citenamefont {Bongs},\ and\ \citenamefont {Lesanovsky}}]{Olmos13}%
  \BibitemOpen
  \bibfield  {author} {\bibinfo {author} {\bibfnamefont {B.}~\bibnamefont
  {Olmos}}, \bibinfo {author} {\bibfnamefont {D.}~\bibnamefont {Yu}}, \bibinfo
  {author} {\bibfnamefont {Y.}~\bibnamefont {Singh}}, \bibinfo {author}
  {\bibfnamefont {F.}~\bibnamefont {Schreck}}, \bibinfo {author} {\bibfnamefont
  {K.}~\bibnamefont {Bongs}},\ and\ \bibinfo {author} {\bibfnamefont
  {I.}~\bibnamefont {Lesanovsky}},\ }\bibfield  {title} {\bibinfo {title}
  {Long-range interacting many-body systems with alkaline-earth-metal atoms},\
  }\href {https://doi.org/10.1103/PhysRevLett.110.143602} {\bibfield  {journal}
  {\bibinfo  {journal} {Phys. Rev. Lett.}\ }\textbf {\bibinfo {volume} {110}},\
  \bibinfo {pages} {143602} (\bibinfo {year} {2013})}\BibitemShut {NoStop}%
\bibitem [{\citenamefont {Jones}\ \emph {et~al.}(2017)\citenamefont {Jones},
  \citenamefont {Saint},\ and\ \citenamefont {Olmos}}]{Olmos16}%
  \BibitemOpen
  \bibfield  {author} {\bibinfo {author} {\bibfnamefont {R.}~\bibnamefont
  {Jones}}, \bibinfo {author} {\bibfnamefont {R.}~\bibnamefont {Saint}},\ and\
  \bibinfo {author} {\bibfnamefont {B.}~\bibnamefont {Olmos}},\ }\bibfield
  {title} {\bibinfo {title} {Far-field resonance fluorescence from a
  dipole-interacting laser-driven cold atomic gas},\ }\href
  {http://stacks.iop.org/0953-4075/50/i=1/a=014004} {\bibfield  {journal}
  {\bibinfo  {journal} {Journal of Physics B: Atomic, Molecular and Optical
  Physics}\ }\textbf {\bibinfo {volume} {50}},\ \bibinfo {pages} {014004}
  (\bibinfo {year} {2017})}\BibitemShut {NoStop}%
\bibitem [{\citenamefont {Williamson}\ \emph {et~al.}(2020)\citenamefont
  {Williamson}, \citenamefont {Borgh},\ and\ \citenamefont
  {Ruostekoski}}]{Williamson2020b}%
  \BibitemOpen
  \bibfield  {author} {\bibinfo {author} {\bibfnamefont {L.~A.}\ \bibnamefont
  {Williamson}}, \bibinfo {author} {\bibfnamefont {M.~O.}\ \bibnamefont
  {Borgh}},\ and\ \bibinfo {author} {\bibfnamefont {J.}~\bibnamefont
  {Ruostekoski}},\ }\bibfield  {title} {\bibinfo {title} {{Superatom Picture of
  Collective Nonclassical Light Emission and Dipole Blockade in Atom Arrays}},\
  }\href {https://doi.org/10.1103/PhysRevLett.125.073602} {\bibfield  {journal}
  {\bibinfo  {journal} {Phys. Rev. Lett.}\ }\textbf {\bibinfo {volume} {125}},\
  \bibinfo {pages} {073602} (\bibinfo {year} {2020})}\BibitemShut {NoStop}%
\bibitem [{\citenamefont {Cidrim}\ \emph {et~al.}(2020)\citenamefont {Cidrim},
  \citenamefont {{do Espirito Santo}}, \citenamefont {Schachenmayer},
  \citenamefont {Kaiser},\ and\ \citenamefont {Bachelard}}]{cidrim2020}%
  \BibitemOpen
  \bibfield  {author} {\bibinfo {author} {\bibfnamefont {A.}~\bibnamefont
  {Cidrim}}, \bibinfo {author} {\bibfnamefont {T.~S.}\ \bibnamefont {{do
  Espirito Santo}}}, \bibinfo {author} {\bibfnamefont {J.}~\bibnamefont
  {Schachenmayer}}, \bibinfo {author} {\bibfnamefont {R.}~\bibnamefont
  {Kaiser}},\ and\ \bibinfo {author} {\bibfnamefont {R.}~\bibnamefont
  {Bachelard}},\ }\bibfield  {title} {\bibinfo {title} {{Photon Blockade with
  Ground-State Neutral Atoms}},\ }\href
  {https://doi.org/10.1103/PhysRevLett.125.073601} {\bibfield  {journal}
  {\bibinfo  {journal} {Phys. Rev. Lett.}\ }\textbf {\bibinfo {volume} {125}},\
  \bibinfo {pages} {073601} (\bibinfo {year} {2020})}\BibitemShut {NoStop}%
\bibitem [{\citenamefont {Parmee}\ and\ \citenamefont
  {Ruostekoski}(2020)}]{Parmee2020}%
  \BibitemOpen
  \bibfield  {author} {\bibinfo {author} {\bibfnamefont {C.~D.}\ \bibnamefont
  {Parmee}}\ and\ \bibinfo {author} {\bibfnamefont {J.}~\bibnamefont
  {Ruostekoski}},\ }\bibfield  {title} {\bibinfo {title} {Signatures of optical
  phase transitions in superradiant and subradiant atomic arrays},\ }\href
  {https://doi.org/10.1038/s42005-020-00476-1} {\bibfield  {journal} {\bibinfo
  {journal} {Communications Physics}\ }\textbf {\bibinfo {volume} {3}},\
  \bibinfo {pages} {205} (\bibinfo {year} {2020})}\BibitemShut {NoStop}%
\bibitem [{\citenamefont {Scarlatella}\ and\ \citenamefont
  {Cooper}(2024)}]{scarlatella2024}%
  \BibitemOpen
  \bibfield  {author} {\bibinfo {author} {\bibfnamefont {O.}~\bibnamefont
  {Scarlatella}}\ and\ \bibinfo {author} {\bibfnamefont {N.~R.}\ \bibnamefont
  {Cooper}},\ }\bibfield  {title} {\bibinfo {title} {Fate of the {Mollow}
  triplet in strongly coupled atomic arrays},\ }\href
  {https://doi.org/10.1103/PhysRevA.110.L041305} {\bibfield  {journal}
  {\bibinfo  {journal} {Phys. Rev. A}\ }\textbf {\bibinfo {volume} {110}},\
  \bibinfo {pages} {L041305} (\bibinfo {year} {2024})}\BibitemShut {NoStop}%
\bibitem [{\citenamefont {Tečer}\ \emph {et~al.}(2025)\citenamefont {Tečer},
  \citenamefont {Calajó},\ and\ \citenamefont {Liberto}}]{tecer2025}%
  \BibitemOpen
  \bibfield  {author} {\bibinfo {author} {\bibfnamefont {M.}~\bibnamefont
  {Tečer}}, \bibinfo {author} {\bibfnamefont {G.}~\bibnamefont {Calajó}},\
  and\ \bibinfo {author} {\bibfnamefont {M.~D.}\ \bibnamefont {Liberto}},\
  }\href {https://arxiv.org/abs/2505.10623} {\bibinfo {title} {Flat band
  mediated photon-photon interactions in 2d waveguide qed networks}} (\bibinfo
  {year} {2025}),\ \Eprint {https://arxiv.org/abs/2505.10623} {arXiv:2505.10623
  [quant-ph]} \BibitemShut {NoStop}%
\bibitem [{\citenamefont {Perczel}\ \emph {et~al.}(2017)\citenamefont
  {Perczel}, \citenamefont {Borregaard}, \citenamefont {Chang}, \citenamefont
  {Pichler}, \citenamefont {Yelin}, \citenamefont {Zoller},\ and\ \citenamefont
  {Lukin}}]{Perczel2017a}%
  \BibitemOpen
  \bibfield  {author} {\bibinfo {author} {\bibfnamefont {J.}~\bibnamefont
  {Perczel}}, \bibinfo {author} {\bibfnamefont {J.}~\bibnamefont {Borregaard}},
  \bibinfo {author} {\bibfnamefont {D.~E.}\ \bibnamefont {Chang}}, \bibinfo
  {author} {\bibfnamefont {H.}~\bibnamefont {Pichler}}, \bibinfo {author}
  {\bibfnamefont {S.~F.}\ \bibnamefont {Yelin}}, \bibinfo {author}
  {\bibfnamefont {P.}~\bibnamefont {Zoller}},\ and\ \bibinfo {author}
  {\bibfnamefont {M.~D.}\ \bibnamefont {Lukin}},\ }\bibfield  {title} {\bibinfo
  {title} {Photonic band structure of two-dimensional atomic lattices},\ }\href
  {https://doi.org/10.1103/PhysRevA.96.063801} {\bibfield  {journal} {\bibinfo
  {journal} {Phys. Rev. A}\ }\textbf {\bibinfo {volume} {96}},\ \bibinfo
  {pages} {063801} (\bibinfo {year} {2017})}\BibitemShut {NoStop}%
\bibitem [{\citenamefont {Bettles}\ \emph {et~al.}(2017)\citenamefont
  {Bettles}, \citenamefont {Min\'a\ifmmode~\check{r}\else \v{r}\fi{}},
  \citenamefont {Adams}, \citenamefont {Lesanovsky},\ and\ \citenamefont
  {Olmos}}]{Bettles_topo}%
  \BibitemOpen
  \bibfield  {author} {\bibinfo {author} {\bibfnamefont {R.~J.}\ \bibnamefont
  {Bettles}}, \bibinfo {author} {\bibfnamefont {J.~c.~v.}\ \bibnamefont
  {Min\'a\ifmmode~\check{r}\else \v{r}\fi{}}}, \bibinfo {author} {\bibfnamefont
  {C.~S.}\ \bibnamefont {Adams}}, \bibinfo {author} {\bibfnamefont
  {I.}~\bibnamefont {Lesanovsky}},\ and\ \bibinfo {author} {\bibfnamefont
  {B.}~\bibnamefont {Olmos}},\ }\bibfield  {title} {\bibinfo {title}
  {Topological properties of a dense atomic lattice gas},\ }\href
  {https://doi.org/10.1103/PhysRevA.96.041603} {\bibfield  {journal} {\bibinfo
  {journal} {Phys. Rev. A}\ }\textbf {\bibinfo {volume} {96}},\ \bibinfo
  {pages} {041603} (\bibinfo {year} {2017})}\BibitemShut {NoStop}%
\bibitem [{\citenamefont {Biehs}\ and\ \citenamefont
  {Agarwal}(2017)}]{Biehs17}%
  \BibitemOpen
  \bibfield  {author} {\bibinfo {author} {\bibfnamefont {S.-A.}\ \bibnamefont
  {Biehs}}\ and\ \bibinfo {author} {\bibfnamefont {G.~S.}\ \bibnamefont
  {Agarwal}},\ }\bibfield  {title} {\bibinfo {title} {Qubit entanglement across
  $\ensuremath{\epsilon}$-near-zero media},\ }\href
  {https://doi.org/10.1103/PhysRevA.96.022308} {\bibfield  {journal} {\bibinfo
  {journal} {Phys. Rev. A}\ }\textbf {\bibinfo {volume} {96}},\ \bibinfo
  {pages} {022308} (\bibinfo {year} {2017})}\BibitemShut {NoStop}%
\bibitem [{\citenamefont {Li}\ \emph {et~al.}(2019)\citenamefont {Li},
  \citenamefont {Nemilentsau},\ and\ \citenamefont {Argyropoulos}}]{li19}%
  \BibitemOpen
  \bibfield  {author} {\bibinfo {author} {\bibfnamefont {Y.}~\bibnamefont
  {Li}}, \bibinfo {author} {\bibfnamefont {A.}~\bibnamefont {Nemilentsau}},\
  and\ \bibinfo {author} {\bibfnamefont {C.}~\bibnamefont {Argyropoulos}},\
  }\bibfield  {title} {\bibinfo {title} {Resonance energy transfer and quantum
  entanglement mediated by epsilon-near-zero and other plasmonic waveguide
  systems},\ }\href {https://doi.org/10.1039/C9NR05083C} {\bibfield  {journal}
  {\bibinfo  {journal} {Nanoscale}\ }\textbf {\bibinfo {volume} {11}},\
  \bibinfo {pages} {14635} (\bibinfo {year} {2019})}\BibitemShut {NoStop}%
\bibitem [{\citenamefont {Chen}\ \emph {et~al.}(2011)\citenamefont {Chen},
  \citenamefont {Wang}, \citenamefont {Jia}, \citenamefont {Geng},
  \citenamefont {Li}, \citenamefont {Feng}, \citenamefont {Qian}, \citenamefont
  {Liang}, \citenamefont {Zhang}, \citenamefont {Gu},\ and\ \citenamefont
  {Zhuang}}]{Chen2011}%
  \BibitemOpen
  \bibfield  {author} {\bibinfo {author} {\bibfnamefont {J.}~\bibnamefont
  {Chen}}, \bibinfo {author} {\bibfnamefont {Y.}~\bibnamefont {Wang}}, \bibinfo
  {author} {\bibfnamefont {B.}~\bibnamefont {Jia}}, \bibinfo {author}
  {\bibfnamefont {T.}~\bibnamefont {Geng}}, \bibinfo {author} {\bibfnamefont
  {X.}~\bibnamefont {Li}}, \bibinfo {author} {\bibfnamefont {L.}~\bibnamefont
  {Feng}}, \bibinfo {author} {\bibfnamefont {W.}~\bibnamefont {Qian}}, \bibinfo
  {author} {\bibfnamefont {B.}~\bibnamefont {Liang}}, \bibinfo {author}
  {\bibfnamefont {X.}~\bibnamefont {Zhang}}, \bibinfo {author} {\bibfnamefont
  {M.}~\bibnamefont {Gu}},\ and\ \bibinfo {author} {\bibfnamefont
  {S.}~\bibnamefont {Zhuang}},\ }\bibfield  {title} {\bibinfo {title}
  {Observation of the inverse doppler effect in negative-index materials at
  optical frequencies},\ }\href {https://doi.org/10.1038/nphoton.2011.17}
  {\bibfield  {journal} {\bibinfo  {journal} {Nature Photonics}\ }\textbf
  {\bibinfo {volume} {5}},\ \bibinfo {pages} {239} (\bibinfo {year}
  {2011})}\BibitemShut {NoStop}%
\bibitem [{\citenamefont {Lobet}\ \emph {et~al.}(2020)\citenamefont {Lobet},
  \citenamefont {Liberal}, \citenamefont {Knall}, \citenamefont {Alam},
  \citenamefont {Reshef}, \citenamefont {Boyd}, \citenamefont {Engheta},\ and\
  \citenamefont {Mazur}}]{Lobet20}%
  \BibitemOpen
  \bibfield  {author} {\bibinfo {author} {\bibfnamefont {M.}~\bibnamefont
  {Lobet}}, \bibinfo {author} {\bibfnamefont {I.}~\bibnamefont {Liberal}},
  \bibinfo {author} {\bibfnamefont {E.~N.}\ \bibnamefont {Knall}}, \bibinfo
  {author} {\bibfnamefont {M.~Z.}\ \bibnamefont {Alam}}, \bibinfo {author}
  {\bibfnamefont {O.}~\bibnamefont {Reshef}}, \bibinfo {author} {\bibfnamefont
  {R.~W.}\ \bibnamefont {Boyd}}, \bibinfo {author} {\bibfnamefont
  {N.}~\bibnamefont {Engheta}},\ and\ \bibinfo {author} {\bibfnamefont
  {E.}~\bibnamefont {Mazur}},\ }\bibfield  {title} {\bibinfo {title}
  {Fundamental radiative processes in near-zero-index media of various
  dimensionalities},\ }\href {https://doi.org/10.1021/acsphotonics.0c00782}
  {\bibfield  {journal} {\bibinfo  {journal} {ACS Photonics}\ }\textbf
  {\bibinfo {volume} {7}},\ \bibinfo {pages} {1965} (\bibinfo {year}
  {2020})}\BibitemShut {NoStop}%
\bibitem [{\citenamefont {Zhou}\ \emph {et~al.}(2020)\citenamefont {Zhou},
  \citenamefont {Alam}, \citenamefont {Karimi}, \citenamefont {Upham},
  \citenamefont {Reshef}, \citenamefont {Liu}, \citenamefont {Willner},\ and\
  \citenamefont {Boyd}}]{Zhou20}%
  \BibitemOpen
  \bibfield  {author} {\bibinfo {author} {\bibfnamefont {Y.}~\bibnamefont
  {Zhou}}, \bibinfo {author} {\bibfnamefont {M.~Z.}\ \bibnamefont {Alam}},
  \bibinfo {author} {\bibfnamefont {M.}~\bibnamefont {Karimi}}, \bibinfo
  {author} {\bibfnamefont {J.}~\bibnamefont {Upham}}, \bibinfo {author}
  {\bibfnamefont {O.}~\bibnamefont {Reshef}}, \bibinfo {author} {\bibfnamefont
  {C.}~\bibnamefont {Liu}}, \bibinfo {author} {\bibfnamefont {A.~E.}\
  \bibnamefont {Willner}},\ and\ \bibinfo {author} {\bibfnamefont {R.~W.}\
  \bibnamefont {Boyd}},\ }\bibfield  {title} {\bibinfo {title} {Broadband
  frequency translation through time refraction in an epsilon-near-zero
  material},\ }\href {https://doi.org/10.1038/s41467-020-15682-2} {\bibfield
  {journal} {\bibinfo  {journal} {Nature Communications}\ }\textbf {\bibinfo
  {volume} {11}},\ \bibinfo {pages} {2180} (\bibinfo {year}
  {2020})}\BibitemShut {NoStop}%
\bibitem [{\citenamefont {V{\'a}zquez-Lozano}\ and\ \citenamefont
  {Liberal}(2023)}]{Lozano23}%
  \BibitemOpen
  \bibfield  {author} {\bibinfo {author} {\bibfnamefont {J.~E.}\ \bibnamefont
  {V{\'a}zquez-Lozano}}\ and\ \bibinfo {author} {\bibfnamefont
  {I.}~\bibnamefont {Liberal}},\ }\bibfield  {title} {\bibinfo {title}
  {Incandescent temporal metamaterials},\ }\href
  {https://doi.org/10.1038/s41467-023-40281-2} {\bibfield  {journal} {\bibinfo
  {journal} {Nature Communications}\ }\textbf {\bibinfo {volume} {14}},\
  \bibinfo {pages} {4606} (\bibinfo {year} {2023})}\BibitemShut {NoStop}%
\bibitem [{\citenamefont {Tirole}\ \emph {et~al.}(2023)\citenamefont {Tirole},
  \citenamefont {Vezzoli}, \citenamefont {Galiffi}, \citenamefont {Robertson},
  \citenamefont {Maurice}, \citenamefont {Tilmann}, \citenamefont {Maier},
  \citenamefont {Pendry},\ and\ \citenamefont {Sapienza}}]{Tirole23}%
  \BibitemOpen
  \bibfield  {author} {\bibinfo {author} {\bibfnamefont {R.}~\bibnamefont
  {Tirole}}, \bibinfo {author} {\bibfnamefont {S.}~\bibnamefont {Vezzoli}},
  \bibinfo {author} {\bibfnamefont {E.}~\bibnamefont {Galiffi}}, \bibinfo
  {author} {\bibfnamefont {I.}~\bibnamefont {Robertson}}, \bibinfo {author}
  {\bibfnamefont {D.}~\bibnamefont {Maurice}}, \bibinfo {author} {\bibfnamefont
  {B.}~\bibnamefont {Tilmann}}, \bibinfo {author} {\bibfnamefont {S.~A.}\
  \bibnamefont {Maier}}, \bibinfo {author} {\bibfnamefont {J.~B.}\ \bibnamefont
  {Pendry}},\ and\ \bibinfo {author} {\bibfnamefont {R.}~\bibnamefont
  {Sapienza}},\ }\bibfield  {title} {\bibinfo {title} {Double-slit time
  diffraction at optical frequencies},\ }\href
  {https://doi.org/10.1038/s41567-023-01993-w} {\bibfield  {journal} {\bibinfo
  {journal} {Nature Physics}\ }\textbf {\bibinfo {volume} {19}},\ \bibinfo
  {pages} {999} (\bibinfo {year} {2023})}\BibitemShut {NoStop}%
\bibitem [{\citenamefont {Benisty}\ \emph {et~al.}(2022)\citenamefont
  {Benisty}, \citenamefont {Greffet},\ and\ \citenamefont
  {Lalanne}}]{Greffet_nanophotonics}%
  \BibitemOpen
  \bibfield  {author} {\bibinfo {author} {\bibfnamefont {H.}~\bibnamefont
  {Benisty}}, \bibinfo {author} {\bibfnamefont {J.-J.}\ \bibnamefont
  {Greffet}},\ and\ \bibinfo {author} {\bibfnamefont {P.}~\bibnamefont
  {Lalanne}},\ }\href@noop {} {\emph {\bibinfo {title} {Introduction to
  Nanophotonics}}},\ \bibinfo {edition} {1st}\ ed.\ (\bibinfo  {publisher}
  {Oxford University Press, Oxford},\ \bibinfo {year} {2022})\BibitemShut
  {NoStop}%
\bibitem [{\citenamefont {Novotny}\ and\ \citenamefont
  {Hecht}(2012)}]{Novotny_nanooptics}%
  \BibitemOpen
  \bibfield  {author} {\bibinfo {author} {\bibfnamefont {L.}~\bibnamefont
  {Novotny}}\ and\ \bibinfo {author} {\bibfnamefont {B.}~\bibnamefont
  {Hecht}},\ }\href@noop {} {\emph {\bibinfo {title} {Principles of
  Nano-Optics}}},\ \bibinfo {edition} {2nd}\ ed.\ (\bibinfo  {publisher}
  {Cambridge University Press, Cambridge},\ \bibinfo {year} {2012})\BibitemShut
  {NoStop}%
\bibitem [{\citenamefont {Belov}\ and\ \citenamefont
  {Simovski}(2005)}]{Belov05}%
  \BibitemOpen
  \bibfield  {author} {\bibinfo {author} {\bibfnamefont {P.~A.}\ \bibnamefont
  {Belov}}\ and\ \bibinfo {author} {\bibfnamefont {C.~R.}\ \bibnamefont
  {Simovski}},\ }\bibfield  {title} {\bibinfo {title} {Homogenization of
  electromagnetic crystals formed by uniaxial resonant scatterers},\ }\href
  {https://doi.org/10.1103/PhysRevE.72.026615} {\bibfield  {journal} {\bibinfo
  {journal} {Phys. Rev. E}\ }\textbf {\bibinfo {volume} {72}},\ \bibinfo
  {pages} {026615} (\bibinfo {year} {2005})}\BibitemShut {NoStop}%
\bibitem [{\citenamefont {Belov}\ and\ \citenamefont
  {Simovski}(2006)}]{belov06}%
  \BibitemOpen
  \bibfield  {author} {\bibinfo {author} {\bibfnamefont {P.~A.}\ \bibnamefont
  {Belov}}\ and\ \bibinfo {author} {\bibfnamefont {C.~R.}\ \bibnamefont
  {Simovski}},\ }\bibfield  {title} {\bibinfo {title} {Boundary conditions for
  interfaces of electromagnetic crystals and the generalized ewald-oseen
  extinction principle},\ }\href {https://doi.org/10.1103/PhysRevB.73.045102}
  {\bibfield  {journal} {\bibinfo  {journal} {Phys. Rev. B}\ }\textbf {\bibinfo
  {volume} {73}},\ \bibinfo {pages} {045102} (\bibinfo {year}
  {2006})}\BibitemShut {NoStop}%
\bibitem [{\citenamefont {Shahmoon}\ \emph {et~al.}(2017)\citenamefont
  {Shahmoon}, \citenamefont {Wild}, \citenamefont {Lukin},\ and\ \citenamefont
  {Yelin}}]{Shahmoon}%
  \BibitemOpen
  \bibfield  {author} {\bibinfo {author} {\bibfnamefont {E.}~\bibnamefont
  {Shahmoon}}, \bibinfo {author} {\bibfnamefont {D.~S.}\ \bibnamefont {Wild}},
  \bibinfo {author} {\bibfnamefont {M.~D.}\ \bibnamefont {Lukin}},\ and\
  \bibinfo {author} {\bibfnamefont {S.~F.}\ \bibnamefont {Yelin}},\ }\bibfield
  {title} {\bibinfo {title} {Cooperative resonances in light scattering from
  two-dimensional atomic arrays},\ }\href
  {https://doi.org/10.1103/PhysRevLett.118.113601} {\bibfield  {journal}
  {\bibinfo  {journal} {Phys. Rev. Lett.}\ }\textbf {\bibinfo {volume} {118}},\
  \bibinfo {pages} {113601} (\bibinfo {year} {2017})}\BibitemShut {NoStop}%
\bibitem [{\citenamefont {Javanainen}\ and\ \citenamefont
  {Rajapakse}(2019)}]{Javanainen19}%
  \BibitemOpen
  \bibfield  {author} {\bibinfo {author} {\bibfnamefont {J.}~\bibnamefont
  {Javanainen}}\ and\ \bibinfo {author} {\bibfnamefont {R.}~\bibnamefont
  {Rajapakse}},\ }\bibfield  {title} {\bibinfo {title} {Light propagation in
  systems involving two-dimensional atomic lattices},\ }\href
  {https://doi.org/10.1103/PhysRevA.100.013616} {\bibfield  {journal} {\bibinfo
   {journal} {Phys. Rev. A}\ }\textbf {\bibinfo {volume} {100}},\ \bibinfo
  {pages} {013616} (\bibinfo {year} {2019})}\BibitemShut {NoStop}%
\end{thebibliography}
\end{document}